\PassOptionsToPackage{table,xcdraw,dvipsnames}{xcolor}
\PassOptionsToPackage{hyphens}{url}
\PassOptionsToPackage{unicode}{hyperref}
\PassOptionsToPackage{naturalnames}{hyperref}

\documentclass[sigplan, screen, authorversion]{acmart}

\usepackage[normalem]{ulem}
\usepackage{enumitem}
\usepackage{multirow} 
\usepackage{adjustbox} % resizing of tables
\usepackage{subcaption} % Recommended in LaTeX http://www.acm.org/binaries/content/assets/publications/consolidated-tex-template/acmart.pdf having 3 figures in a row
\usepackage{tabularx} % INCLUDED in the acm package having 3 figures in a row
\usepackage{tikz} %Create PostScript and PDF graphics
\usepackage{pifont}
\newcommand{\cmark}{\ding{51}}%

\hypersetup{
    colorlinks=true,
    pdfborder={0 0 0},
}
\usepackage{listings}

\newif\ifshowcomment
\showcommentfalse

\ifshowcomment

\newcommand{\todo}[1]{\noindent\textsf{\color{orange}{[{ToDo: \it #1}]}}}
\newcommand{\newtext}[1]{\textcolor{blue}{#1}} % Added new texts
\newcommand{\snewtext}[1]{#1} 
\newcommand{\boris}[1]{\noindent\textsf{\color{Violet}{[Boris: {\it#1}]}}}

\newcommand{\yijun}[1]{\noindent\textsf{\color{OliveGreen}{[Yijun: {\it#1}]}}}
\newcommand{\antonis}[1]{\noindent\textsf{\color{magenta}{[Antonis: {\it#1}]}}}
\newcommand{\zhaowei}[1]{\noindent\textsf{\color{brown}{[Zhaowei: {\it#1}]}}}

\else
\newcommand{\newtext}[1]{#1} 
\newcommand{\snewtext}[1]{#1} 

\newcommand{\todo}[1]{}
\newcommand{\antonis}[1]{}
\newcommand{\boris}[1]{}
\newcommand{\vijay}[1]{}
\newcommand{\arpit}[1]{}
\newcommand{\yijun}[1]{}
\newcommand{\zhaowei}[1]{}

\fi

\newif\ifshowirl
\showirlfalse

\ifshowirl

\newcommand{\irl}[1]{\textcolor{red}{#1}}  % Modified texts

\else
\newcommand{\irl}[1]{}
\fi

\newcommand\blfootnote[1]{%
  \begingroup
  \renewcommand\thefootnote{}\footnote{#1}%
  \addtocounter{footnote}{-1}%
  \endgroup
}

\newcommand*\circled[1]{\tikz[baseline=(char.base)]{
            \node[shape=circle, text=white, fill=black, draw,inner sep=0.5pt] (char) {#1};}}

\newcommand{\beginbsec}[1]{\noindent\textbf{#1.}}

\newcommand{\mylisting}[1]{\scalebox{0.9}{\texttt{#1}}}

\newcommand{\CAP}[1]{\scalebox{0.85}{#1}}

\newcommand{\squishlist}{
 \begin{list}{$\bullet$}
  { \setlength{\itemsep}{2pt}
     \setlength{\parsep}{0pt}
     \setlength{\topsep}{2pt}
     \setlength{\partopsep}{0pt}
     \setlength{\leftmargin}{1em}
     \setlength{\labelwidth}{1em}
     \setlength{\labelsep}{0.5em} } 
}

\newcommand{\squishlistContrib}{ %
 \begin{list}{$\bullet$}
  { \setlength{\itemsep}{2pt}
     \setlength{\parsep}{0pt}
     \setlength{\topsep}{2pt}
     \setlength{\partopsep}{0pt}
     \setlength{\leftmargin}{1em}
     \setlength{\labelwidth}{1em}
     \setlength{\labelsep}{0.5em} }
}
\newcommand{\squishend}{ \end{list}  }

\newcommand{\squishenum}{\begin{enumerate}[itemsep=0.5pt,parsep=0pt,topsep=0pt,partopsep=0pt,leftmargin=1.5em,labelwidth=1em,labelsep=0.5em]{}}
\newcommand{\squishenumend}{\end{enumerate}}

\def\eg{e.g.,~} % ``for example''
\def\INVNOSPACE{\CAP{R-INV}} 
\def\ACKNOSPACE{\CAP{R-ACK}}
\def\VALNOSPACE{\CAP{R-VAL}}
\def\INV{{\INVNOSPACE} } 
\def\ACK{{\ACKNOSPACE} }
\def\VAL{{\VALNOSPACE} }

\begin{document}

\title{Zeus: Locality-aware Distributed Transactions}

\acmYear{2021}\copyrightyear{2021}
\setcopyright{acmlicensed}
\acmConference[EuroSys '21]{Sixteenth European Conference on Computer Systems}{April 26--29, 2021}{Online, United Kingdom}
\acmBooktitle{Sixteenth European Conference on Computer Systems (EuroSys '21), April 26--29, 2021, Online, United Kingdom}
\acmPrice{15.00}
\acmDOI{10.1145/3447786.3456234}
\acmISBN{978-1-4503-8334-9/21/04}

%% Authors
\author{Antonios Katsarakis\texorpdfstring{$^{\dagger^{*}}$}{}, Yijun Ma\texorpdfstring{$^{\ddagger}$}{},  Zhaowei Tan\texorpdfstring{$^{\mathsection^{*}}$}{}, Andrew Bainbridge, Matthew Balkwill,}
\author{Aleksandar Dragojevic, Boris Grot\texorpdfstring{$^{\dagger}$}{}, Bozidar Radunovic, Yongguang Zhang}

\affiliation{
\institution{
 \texorpdfstring{$^{\dagger}$}{}University of Edinburgh,
 \texorpdfstring{$^{\ddagger}$}{}Fudan University,
 \texorpdfstring{$^{\mathsection}$}{}UCLA,
 \hspace{1pt} Microsoft Research}
}

%% Used for header
\renewcommand{\shortauthors}{A. Katsarakis, et al.}



\begin{abstract}

State-of-the-art distributed in-memory datastores (FaRM, FaSST, DrTM) provide strongly-consistent distributed transactions with high performance and availability. Transactions in those systems are fully general; they can atomically manipulate any set of objects in the store, regardless of their location. 
To achieve this, these systems use complex distributed transactional protocols. Meanwhile, many workloads have a high degree of locality. For such workloads, distributed transactions are an overkill as most operations only access objects located on the same server -- if sharded appropriately.

In this paper, we show that for these workloads, a single-node transactional protocol combined with dynamic object re-sharding and asynchronously pipelined replication can provide the same level of generality with better performance, simpler protocols, and lower developer effort. 
We present Zeus, an in-memory distributed datastore that provides general transactions by acquiring all objects involved in the transaction to the same server and executing a single-node transaction on them. Zeus is fault-tolerant and strongly-consistent. 
At the heart of Zeus is a reliable dynamic object sharding protocol that can move 250K objects per second per server,
allowing Zeus to process millions of transactions per second and outperform more traditional distributed transactions on a wide range of workloads that exhibit locality.
\end{abstract}

\maketitle

\blfootnote{$^{*}$Part of this work was done when the author was in Microsoft Research.}
\vspace{-26.5pt}
\section{Introduction}
\label{sec:introduction}

Cloud applications over commodity infrastructure are becoming increasingly popular. 
They require distributed, fast and reliable datastores.
Recent in-memory datastores that operate within a datacenter and leverage replication for fault-tolerance (FaRM \cite{dragojevic2014farm}, FaSST~\cite{kalia2016fasst}, DrTM~\cite{wei2015fast}) offer strongly-consistent distributed transactions in the order of millions per second.
They do not make any assumptions about the workloads and rely on highly-optimized remote access primitives (e.g., \CAP{RDMA}) to enable a variety of use cases.  

These datastores run \CAP{OLTP} workloads with transactions involving a small number of objects. 
In addition, many applications have a high degree of locality. 
For example, many transactions in a cellular control plane involve one user always accessing the same set of objects (e.g., the nearest base station, or the same call forwarding number~\cite{TATP:2009}). 
Many Internet middle-boxes mostly access the same state for all packets of a single flow (e.g., intrusion detection systems~\cite{Woo:18}).
Bank transactions often recur between the same parties~\cite{cahill2009serializable, Venmo17, Venmo20}. 
\newtext{As Stonebraker \textit{et al.} report~\cite{Harding:17}, a transactional concurrency control scheme can derive significant benefit from leveraging application specific characteristics such as locality.}

Existing works~\cite{kalia2016fasst, dragojevic2014farm, wei2015fast} can exploit locality through \textit{static sharding} -- iff all objects involved in each transaction are stored on the same node\footnote{\newtext{Throughout the paper we use the terms \textit{node} and \textit{server} interchangeably.}}. 
Consequently, static sharding only helps if the optimal placement is known a priori and never changes. 
However, this is often not the case for two main reasons. 
Firstly, the set of objects involved in a transaction may change over time. 
For instance, as a mobile phone user moves, her {\em cellular handover} transaction involves different base stations.
Secondly, the popularity of each object changes in time, be it a network service or a financial stock. 
If several popular objects are located on the same server, the server becomes a bottleneck, and the popular objects should be spread across servers.
In both cases, {\em the rate of changes in access locality is multiple orders of magnitude lower than the rate of processed transactions} (which is in millions per second). 
We describe these cases in more detail in Section~\ref{sec:background}.

In contrast, \textit{dynamic sharding}, where objects are moved on-demand across nodes, helps both
when the set of objects involved in a transaction changes or when object popularity shifts. 
In the first case, dynamic sharding ensures that all objects involved in a transaction are colocated, thus reducing expensive remote accesses. 
In the second case, dynamic sharding allows to quickly spread out the heavy-hitters, thus alleviating the bottlenecks. 
However, state-of-the-art works~\cite{kalia2016fasst, dragojevic2014farm, wei2015fast} do not support dynamic object sharding.
Once the existing sharding is no longer optimal, they revert to remote transactions that are inherently slower. Remote transactions are slow because they impose the overhead of several round-trips both to execute a transaction via remote accesses and to atomically commit it.
The source of the latter is the complexity of distributed atomic commit for conflict resolution under the uncertainty of faults.

Several systems propose application-level load balancer designs that let applications make a fine-grained decision regarding which node each transaction should be routed to~\cite{adya2016slicer, banerjee2015scaling, ahmad2020low, annamalai2018sharding}.
However, most of these systems rely on custom datastores that either do not provide strong consistency or are not as fast as the state-of-the-art datastores~\cite{kalia2016fasst, dragojevic2014farm, wei2015fast}. 
As argued by Adya \textit{et al.}~\cite{adya2019fast}, there is a need for a general distributed protocol that would provide strongly-consistent transactions and better exploit dynamic locality.

In this paper, we address the problem of high-performance dynamic sharding for transactional workloads by presenting a novel distributed datastore called {\em Zeus}.
The key insight behind Zeus is that, for many workloads, the benefits of local execution outweigh the cost of (relatively infrequent) re-sharding. Zeus capitalizes on this insight through two novel 
reliable 
protocols designed from the ground-up to exploit locality in transactional workloads.
One protocol is responsible for reliable (atomic and fault-tolerant) object ownership migration 
requiring at most 1.5 round-trips during common operation.
Using this protocol, while executing a transaction, Zeus moves all objects to the server executing it and ensures exclusive write access. 
Once that is done, and unless the access pattern changes, all subsequent transactions to this set of objects will be executed entirely locally and eschew the need for a costly distributed conflict resolution. 
The second protocol is a fast reliable commit protocol for the replication of localized transactions. 
By combining these two protocols, Zeus achieves performance and simplicity of single-node transactions with generality of distributed transactions.
\newtext{To further exploit locality, Zeus' reliable commit enables local yet consistent read-only transactions from all replicas.}

Zeus design provides an extra benefit in that it allows easy portability of existing applications. 
Since most Zeus transactions are local, Zeus can pipeline executions without compromising correctness.
A subsequent transaction does not need to wait for the replication of the current one.
This is in contrast to the existing in-memory distributed transactional datastores~\cite{kalia2016fasst, dragojevic2014farm, wei2015fast}, in which each transaction blocks until the replication is finished.
To mitigate the effects of blocking, these datastores use custom user-mode threading that requires substantial effort to port existing applications onto. 
In contrast, Zeus transaction pipelining allows easy porting of legacy applications onto it, making them distributed and reliable while reaping the performance benefits of locality with minimal developer effort.

We implement Zeus and evaluate it on several relevant benchmarks: Smallbank~\cite{cahill2009serializable}, Voter~\cite{oltp2013bench}, \CAP{TATP}~\cite{TATP:2009}. 
We also introduce and implement a new benchmark which models handovers in a cellular network based on observed human mobility patterns. 
To demonstrate the ease of porting existing applications to Zeus, we port several networking applications that exhibit locality: cellular packet gateway~\cite{OpenEPC}, Nginx~\cite{nginx_session_persisten} and \CAP{SCTP} transport protocol~\cite{usrsctp:2015}.

\noindent In brief, the main contributions of this work are as follows:

\squishlist
    \item \scalebox{0.965}{\textbf{Proposes \textit{Zeus}, a reliable locality-aware transactional}}
    \textbf{datastore} (\S\ref{sec:design}) that replicates data in-memory for availability.
    \newtext{Unlike state-of-the-art strongly-consistent transactional datastores, Zeus transactions are fast by virtue of exploiting dynamic sharding and locality that exists in certain transactional workloads (as demonstrated in \S\ref{sec:evaluation}).}
    
    \item \textbf{Introduces two reliable protocols} (\S\ref{sec:sharding-protocol}, \ref{sec:zeus-protocol}). An \textit{ownership protocol} for dynamic sharding that quickly alters object placement and access levels across replicas; and a transactional protocol for fast pipelined \newtext{\textit{reliable commit} and local read-only transactions from all replicas}. Both protocols, which ensure the strongest consistency under concurrency and faults, are formally verified in \CAP{TLA$^{+}$}. 
    
    \item \textbf{Implements and evaluates Zeus} (\S\ref{sec:system}, \ref{sec:evaluation}) 
    over \CAP{DPDK} on a six node cluster, using three standard \CAP{OLTP} benchmarks and a new cellular handover benchmark. For workloads with high access locality, 
    % Zeus performs up to 2$\times$ better than 
    Zeus achieves up to 2$\times$ the performance of 
    state-of-the-art \CAP{RDMA}-optimized systems, while using less network bandwidth and without relying on \CAP{RDMA}.
    On the handovers benchmark Zeus performance with dynamic sharding is just 4\% to 9\% from the ideal of all local accesses.
    It also shows the ease of portability by porting three legacy applications showing scalability and reliability with little or no performance drop. 
\squishend
\section{Objectives and motivation}
\label{sec:background}

We first describe high-level objectives that a data center operator and an application developer desire in a datastore. 
We next discuss the opportunities that arise with local access patterns and why they have not been explored fully before. 

\subsection{Datastore design objectives}
\label{sec:design_objectives}

Our goal is to design an intra-datacenter \newtext{shared-nothing transactional database} for \CAP{OLTP} workloads that allows programmers to deploy their software on top of a distributed infrastructure without needing to re-architect the application. 
More specifically, we want to provide the following:

\vskip 2pt \noindent {\bf Performance and reliability. } 
\newtext{
Our target is to have a reliable datastore that can process millions of operations per second. 
Furthermore, to 
remain available despite node failures,
% facilitate recovery in case of failure, 
each state update must be replicated across nodes.}

\vskip 2pt \noindent {\bf Transactions. } 
A single operation may arbitrarily access or modify multiple objects. 
A notion of transaction guarantees that either all modifications are committed, or none. 
% An operation sequence may arbitrarily access or modify multiple objects. 
% A transaction over that sequence guarantees that either all modifications are committed, or none. 
This is in contrast to many widely used in-memory key-value stores (e.g.,~\cite{redis:2020}) that essentially provide only single-object atomic abstractions and some generalizations as an afterthought.

\vskip 2pt \noindent {\bf Strong consistency. } 
We want to provide a simple programming model where a programmer has the intuitive notion of a single-copy of state, despite the state being replicated for reliability.
This model requires strongly-consistent distributed transactions guaranteeing strict serializability~\cite{sethi1982}. 
Informally, with strict serializability all transactions appear as if they are atomically performed at a single point in real-time to all replicas in-between their invocation and response.

\vskip 2pt \noindent {\bf Support for legacy applications. } 
State-of-the-art in-mem\-ory datastores~\cite{kalia2016fasst, dragojevic2014farm,wei2015fast} meet the above criteria. 
However, when executing remote transactions, they block the associated threads.  
To mask the performance cost of blocking, they rely on 
transaction multiplexing and 
user-mode threads~\cite{kalia2016fasst}.
% (e.g., Boost user-threads in~\cite{kalia2016fasst}).
However, this makes porting existing applications on top of these frameworks difficult.
Our goal is to provide a datastore that allows legacy applications to run on top of it without 
\newtext{mandating modifications to the existing architecture}.

\subsection{A case for access locality}

As noted in Section~\ref{sec:introduction}, many real-world applications exhibit transactional access patterns with a high degree of locality. 
In these cases, data is usually sharded for efficiency. 
However, the optimal sharding may change in time for two reasons.
One is due to changes in object popularity and the other one due to changes in access locality. 
We use the term \textit{locality} 
to refer to
the temporal reuse of transactions between (spatially related) objects that reside on the same node.

Let us consider changes in locality via an example of call handovers in a cellular network. 
Every time a phone wakes up to process data traffic (a {\em service request}) or goes to sleep (a {\em release request}), the cellular control plane updates various objects related to the phone and to the base station this phone is attached to. 
This is an example of data access locality, where each consecutive operation on the same phone accesses the same two objects (the phone and the base station contexts).

However, the access locality may slowly and gradually change in time due to mobility. 
Every time a cellular user moves from one base station to another, her phone performs a {\em handover} operation.
This is a transaction that involves three entities, the phone, the old base station the user is leaving, and the new base station the user is connecting to. 
As the user travels (e.g. during a daily commute), her phone will perform many such transactions, each involving one object that stays the same (the phone context) and two other objects that continuously change (contexts of the base stations on the way). 
Once the user finishes the commute, the access locality will resume, and every subsequent {\em service request} and {\em release} for the user will again involve a single base station (the one the user is currently attached to, which is different from the one at the beginning of the commute).

This change is slow in time. 
People are stationary most of the time. 
A study~\cite{bostom2013mobility} shows that an average person makes five one-way trips per day with a total length of 100km for drivers and 20km for non-drivers (on average). 
Consequently, handover requests are only between 2.5\% and 5\% of service and release requests~\cite{mohammadkhan2016considerations, cell_params}, while the vast majority of service and release requests repeatedly include the same base station. 
\newtext{Another fact that further improves locality in this case is}
that a base station will only take part in handovers with other base stations that are geographically close to it.

The optimal sharding should adapt to keep the relevant objects together in the same node. 
In this particular example, it should strive to keep the context of a phone and of the base station it is associated to on the same node. 
However, based on the above observations regarding user mobility, 
re-sharding will occasionally need to happen, though only for a single-digit fraction of transactions.
We further discuss and evaluate this example in Section~\ref{sec:evaluation}.

Another example of access locality are peer-to-peer financial transactions. 
Several studies of the popular peer-to-peer mobile payment system Venmo~\cite{Venmo17, Venmo20} show that the transactions mainly occur among groups of friends, and that the transaction graph exhibits a higher local clustering than Facebook and Twitter graphs. 
Moreover, as noted by \newtext{Unger~\textit{et al.}}~\cite{Venmo20}, the network remains largely consistent across the studies, indicating slow temporal change in the interaction graph. 
\newtext{We study this case using publicly available data from a recent Venmo study~\cite{venmo-dataset} and evaluate it on a popular financial transactions benchmark Smallbank~\cite{cahill2009serializable} in Section~\ref{sec:evaluation}}.

The optimal sharding may also change due to a shift in object popularity.
One example of this can be found in the Voter benchmark~\cite{oltp2013bench}, which we evaluate in Section~\ref{sec:evaluation}. 
In a long-lasting online public contest (e.g., Eurovision), many users vote for a few contestants. 
The optimal sharding should spread the load evenly, and would ideally put the most popular contestants each on a separate server,
while potentially grouping the least popular contestants together on a single server. 
However, the popularity of each contestant changes in time, and as she gets more or fewer votes, the optimal sharding changes as well.
As in the previous example, each transaction involves only a few objects (a voter and a contestant) and the frequency of change in the optimal sharding is much lower than the frequency of the voting transactions. 

Another example is stock exchange. 
Between 40–60\% of the volume on the New York Stock Exchange occurs on just 40 out of 4000 stocks~\cite{taft2014store}.
Stock popularity changes at the granularity of hours or days, whereas daily trading volume is on the order of 5-10 billion shares~\cite{stocks}. Thus, while transaction volume is high, the change in popularity is slow. 
Similar to the handover case, the re-sharding will need to happen, but relatively infrequently. 

Existing works~\cite{taft2014store, serafini2016clay, curino2010schism, Rocksteady} propose dynamic sharding to adapt to these kinds of changes. 
\newtext{However, their datastore designs that support re-sharding and provide strong consistency} operate at a sub-Mtps throughput. For instance, Squall~\cite{elmore2015squall} and Rococo~\cite{Rococo} report up to 100 Ktps per server, and Rocksteady~\cite{Rocksteady} up to 700 Ktps per server. 

Meanwhile, state-of-the-art reliable in-memory datastores (e.g., FaRM, FaSST) reach millions of tps per node but have limited support for changes in locality. 
For instance, FaRM only supports static location hints. 
If the access locality changes, both FaRM and FaSST must execute remote transactions. 
Some domain-specific datastores have been built that exploit locality, but they do not meet all design objectives. \newtext{For example, S6~\cite{Woo:18} does not offer replication (a must for availability), while \CAP{FTMB}~\cite{Sherry:15} runs only on one node and replicates on non-volatile storage}. 
Overall, to the best of our knowledge, there is no in-memory datastore that meets all our design objectives and effectively exploits locality. 

\section{Design overview}
\label{sec:design}

We start this section by outlining the Zeus datastore system architecture. 
We then present a high-level overview of the key part of Zeus --- a pair of protocols that exploit locality for high-performance transaction processing with fault-tolerance, strong consistency and programmability.

\subsection{Zeus system architecture}
Zeus exploits request locality and uses an application-level load balancer to enforce it. 
External requests issued to Zeus are issued through a load balancer. 
The load balancer can extract the application level information, locate relevant object keys and always forwards requests with the same set of keys to the same server.
Application-level load balancers are not a new concept. 
Several previous systems have demonstrated such load balancers~\cite{adya2016slicer, Nguyen:2018, ahmad2020low, annamalai2018sharding}. 
We implement a simple one using a distributed, replicated key-value store based on Hermes~\cite{katsarakis2020hermes}. 
We extract a key from each request and look it up in the key-value store. 
If not found, we pick a destination Zeus node at random, store it in the load balancer's key-value store and forward the request.
If the key is found, we forward the request to the corresponding destination. 

Zeus considers a non-byzantine partially synchronous model~\cite{Dwork:1988} with crash-stop node failures and network faults including message reordering, duplication and loss.
It implements a reliable messaging protocol with low-level retransmission to recover lost messages. 
Zeus uses a reliable membership with leases to deal with the uncertainty of detecting node failures. 
Each membership update is tagged with a monotonically increasing epoch id ($e\_id$) and is performed across the deployment only after all node leases have expired. 
This provides the same consistent views of \textit{live nodes} across the deployment despite unreliable failure detection (similar to Zookeeper~\cite{Zookeeper} with leases).
For data reliability, Zeus maintains replicas of each object. 
The replication degree is configurable; however, the higher the degree of replication, the greater the \CAP{CPU} and network overhead, and the lower is the throughput of transactions that modify the state.

\subsection{Zeus protocols overview}
\label{sec:protocol}

Zeus is efficient in executing distributed transactions by forcing them to become local. 
At the heart of Zeus are two separate, loosely connected reliable protocols.
One of them is the \newtext{{\em ownership protocol} responsible for the on-demand migration of the object data from one server to another and changing the access rights (read or write) of servers storing the replica of an object}.
The other one is the {\em reliable commit protocol} for committing the updates performed during a transaction to the replicas. 
As these two protocols are only loosely connected, they can be optimized, verified and tested independently. 

Zeus, \newtext{inspired by hardware transactional memory~\cite{htm}}, executes and commits each transaction locally, on a server designated to be the {\em coordinator} for that transaction. 
\newtext{While} executing a transaction, the coordinator has to secure the appropriate ownership level for each object involved in the transaction.
This is the task of the ownership protocol.
Once the coordinator acquires the required ownership levels \newtext{and finishes execution, it commits the transaction locally. Subsequently, it copies} 
the state of modified objects to backup servers, also called {\em followers}.
The latter is the task of the reliable commit protocol. Crucially, the ownership protocol is invoked only the first time a node accesses an object.
Subsequent transactions proceed without involving it, until another node takes over the ownership  (i.e., locality changes).

\begin{figure}[t]
  \centering
  \includegraphics[width=1.05\columnwidth]{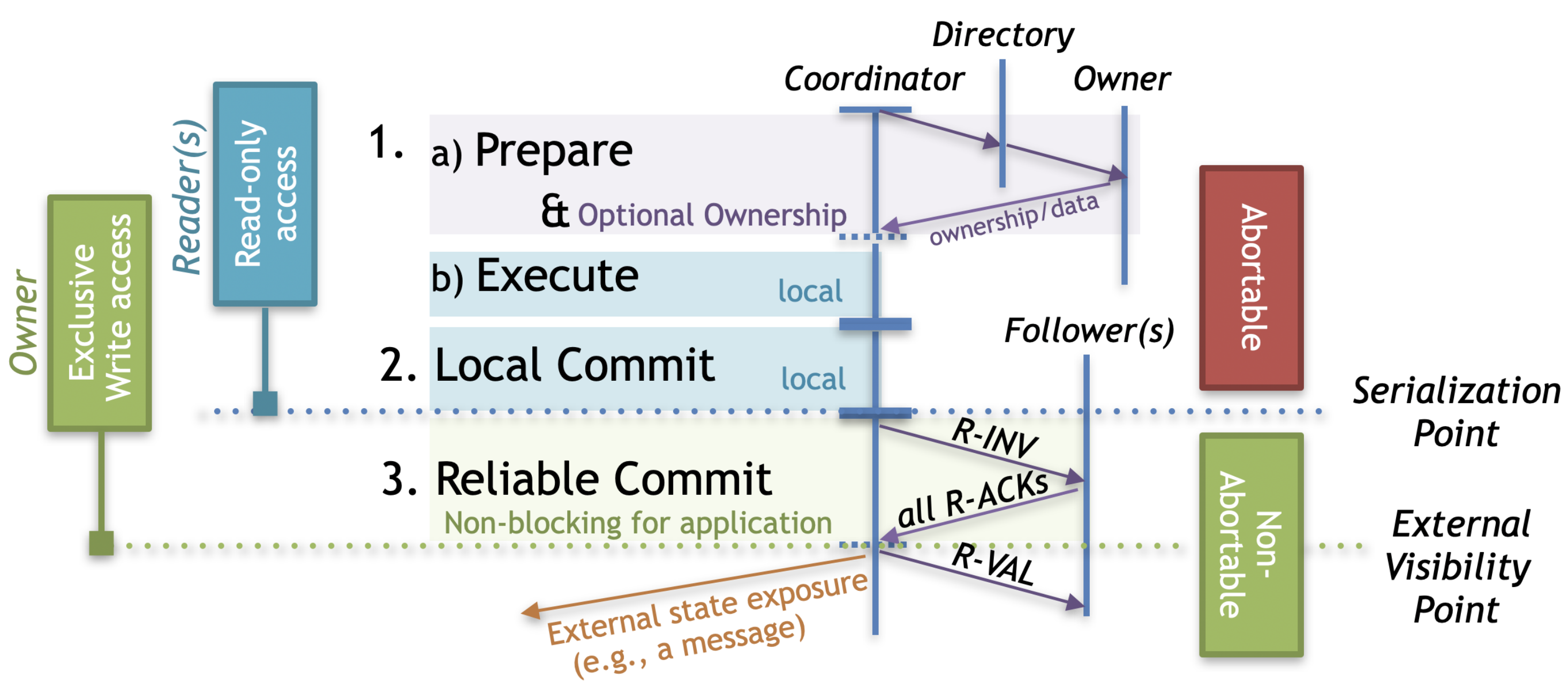}
  \vspace{-20pt}
  \caption{Zeus' locality-aware distributed transactions.}
  \vspace{-5pt}
  \label{fig:zeus}
\end{figure}

At a high level, a transaction in Zeus is carried out through the following three steps (also shown in Figure~\ref{fig:zeus}):

\squishenum
\item \textbf{Prepare \& Execute}: 
\newtext{While the coordinator executes a transaction, prior to accessing an object, it verifies that it holds the appropriate ownership level (read or write) for that object.
If not, it acquires the appropriate ownership level via the {\em ownership protocol} (described in Section~\ref{sec:sharding-protocol}) and continues execution.  
Before performing its first update to an object,
the coordinator creates a private (to the transaction) copy of the object. This private copy is then used for all accesses of the transaction to the object.}

\item \textbf{Local Commit}: The coordinator tries to serialize the transaction locally via a traditional single-node commit.
This commit is local and unreliable but it does not expose any updated values yet to other servers.
We implement a simple multi-threaded local commit that resolves contention across threads using a simplified, local version of the ownership protocol (details in Section~\ref{sec:system}). 

\item \textbf{Reliable Commit}: If the local commit is successful, the coordinator pushes all updates to the followers for data reliability.
In case the coordinator fails in the middle of this process,
the followers recover by safely replaying any pending reliable commit of the failed coordinator.
Both backup and recovery actions are performed by the {\em reliable commit protocol} (details in Section~\ref{sec:zeus-protocol}). 
\squishenumend

Zeus allows only a single server to modify an object at any time. 
This server is called the {\em owner} and is the only node able to use the object to execute write transactions (transactions modifying at least one object).
Each object is replicated on one or more backup servers. 
These backups are active and are called the {\em readers} of the object;
they can perform read-only transactions but not write transactions using the object\footnote{
  Note that a {\em reader} is per object, whereas a {\em follower} is per transaction (potentially spanning multiple objects).
}.
Only the owner and the readers store the content of the object. 
The owner (as a coordinator of write transactions) updates all readers during the reliable commit phase. 
A user can specify and dynamically change the number of readers (i.e., replicas) of each object, making a trade-off between reliability and replication overhead.

\newtext{
Zeus avoids the conventional distributed commit protocols~\cite{Mohan:86, Skeen:1981} which are complex~\cite{Binning:16} because they need to deal with 
distributed conflict resolution and 
the uncertainty of commit or abort after faults.} 
Zeus sidesteps these challenges through a simple invariant that an initiated reliable commit \newtext{is idempotent and cannot be aborted by remote participants}. 
This is accomplished via the exclusive write access of the coordinator and 
the use of {\em\newtext{idempotent} invalidations} (\S~\ref{sec:reliable_commit_protocol}), which are sent to all of the remote participants at the start of the reliable commit. 
In case of a fault, any of the participants can replay the invalidation message which contains enough data to finish the transaction. 

Zeus further introduces two key optimizations.
Firstly, it supports efficient strictly serializable read-only transactions.
Any node that is a \textit{reader} of all objects involved in a read-only transaction is able to execute it without invoking the ownership protocol. 
A read-only transaction does not require a reliable commit phase; as such, it is light-weight and incurs no network traffic. 
Consistency of read-only transactions is enforced through invalidation messages, as a read-only transaction cannot execute on an object that is invalidated. 

\begin{figure}[t]
  \centering
  \includegraphics[width=0.95\columnwidth]{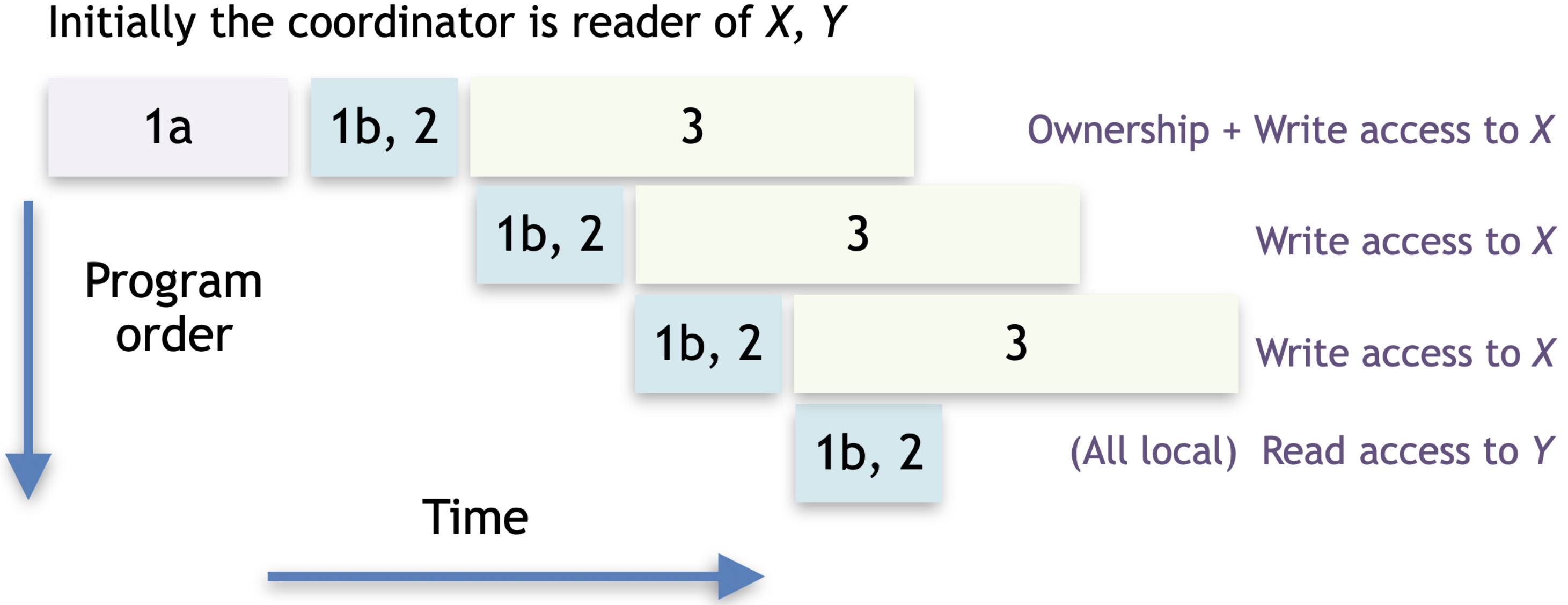}
  \vspace{-10pt}
  \caption{Zeus’ pipelined execution of transactions for objects X and Y, on the same coordinator (labels in boxes are the same as in Figure~\ref{fig:zeus}).}
  \vspace{-5pt}
  \label{fig:zeus_pipe}
\end{figure}

Secondly, a transaction coordinator in Zeus \newtext{pipelines local execution and commit with the reliable commit}, as shown in Figure~\ref{fig:zeus_pipe}.
This is possible because no other server can update the objects at the same time. 
This is guaranteed by the ownership protocol, which ensures that only one node (the current owner) may modify an object.
It is thus safe for the coordinator to keep modifying the same object without waiting for the reliable commit to finish. 
As a consequence, any local transactions to objects for which permissions have already been acquired will not block the application execution. 

We also note that we made a conscious design trade-off to make the ownership protocol blocking, to simplify application portability and to make Zeus transactions (the most frequent operations) non-blocking. 
This means that the application thread stalls when executing an ownership request (phase 1(a) in Figure~\ref{fig:zeus}). 
Such a design is justified because ownership requests are much less frequent than transactions, as discussed in Section~\ref{sec:background}. 
It would be straightforward to improve the performance of the ownership protocol, e.g., via a user-mode thread scheduling framework as in~\cite{kalia2016fasst}; however, that would increase the burden on the developer and likely require re-architecting the application, thus invalidating a key design requirement as laid out in Section~\ref{sec:background}. 

% Finally, we specified both parts of the Zeus protocol in \CAP{TLA$^{+}$} and model checked it. The details are in Section~\ref{sec:evaluation}.
Finally, we specified Zeus' ownership and reliable commit in \CAP{TLA$^{+}$} and model checked them. The details are in Section~\ref{sec:evaluation}.

\section{Reliable ownership}
\label{sec:sharding-protocol}

The reliable ownership 
atomically alters object access rights and transfers content between nodes. 
We start by introducing the main terminology used in the protocol. 
We then overview its operation without faults and contention, and follow by discussing these other cases. 

\vskip 6pt \noindent{\bf Access levels, directory and metadata.}
A node can be the {\em owner}, a {\em reader} or a \newtext{{\em non-replica}} of an object. 
Each object has at most one owner at any time that has an exclusive write and (non-exclusive) read access to it. 
An object can also have several other readers with read access. 
Both the owner and the readers store a replica of the object. 
A \newtext{non-replica} node has neither the access rights nor the data for the object. 

Zeus maintains an {\em ownership directory} where it stores ownership metadata about each object. 
This directory is replicated across three nodes for reliability (even if a Zeus deployment has more nodes). 
The nodes that store directory information are called the {\em directory} nodes. 

\noindent  The directory stores the following metadata for an object: 
\squishlist
\item $o\_state$: the ownership state of the object, which can be \textit{Valid}, \textit{Invalid}, \textit{Request} or \textit{Drive}; 
\item $o\_ts =$<$obj\_ver, node\_id$>: ownership timestamp comprising a monotonically increasing number and a node id; 
\item $o\_replicas$: denotes all nodes storing a  replica of the object and their access rights (i.e., the owner and readers). 
\squishend
These ownership metadata are also stored by each object's owner node. 
The summary of the above is given in Table~\ref{tab:shard_metadata}.

\begin{table}[t!]
\resizebox{\columnwidth}{!}{% use resizebox with textwidth
\begin{tabular}{l|cccc}
\multicolumn{1}{c|}{} & \multicolumn{1}{l}{\textbf{directory}} & \multicolumn{1}{l}{\textbf{owner}} & \multicolumn{1}{l}{\textbf{reader(s)}} & \multicolumn{1}{l}{\textbf{non-replica}} \\
\cline{2-5} 
data               &        & \cmark & \cmark &  \\
ownership metadata  & \cmark & \cmark &        &  \\
ownership levels & - & \textbf{w}/\textbf{r}    & \textbf{r}      & -                             
\end{tabular}
}
\vspace{1pt}
\caption{Data and metadata stored by each node along with their read (r) and exclusive write (w) access permissions.}
\vspace{-17pt}
\label{tab:shard_metadata}
\end{table}

\subsection{Reliable ownership protocol}
\begin{figure}[t]
  \vspace{-3pt}
  \hspace{-13pt}
  \includegraphics[width=1.05\columnwidth]{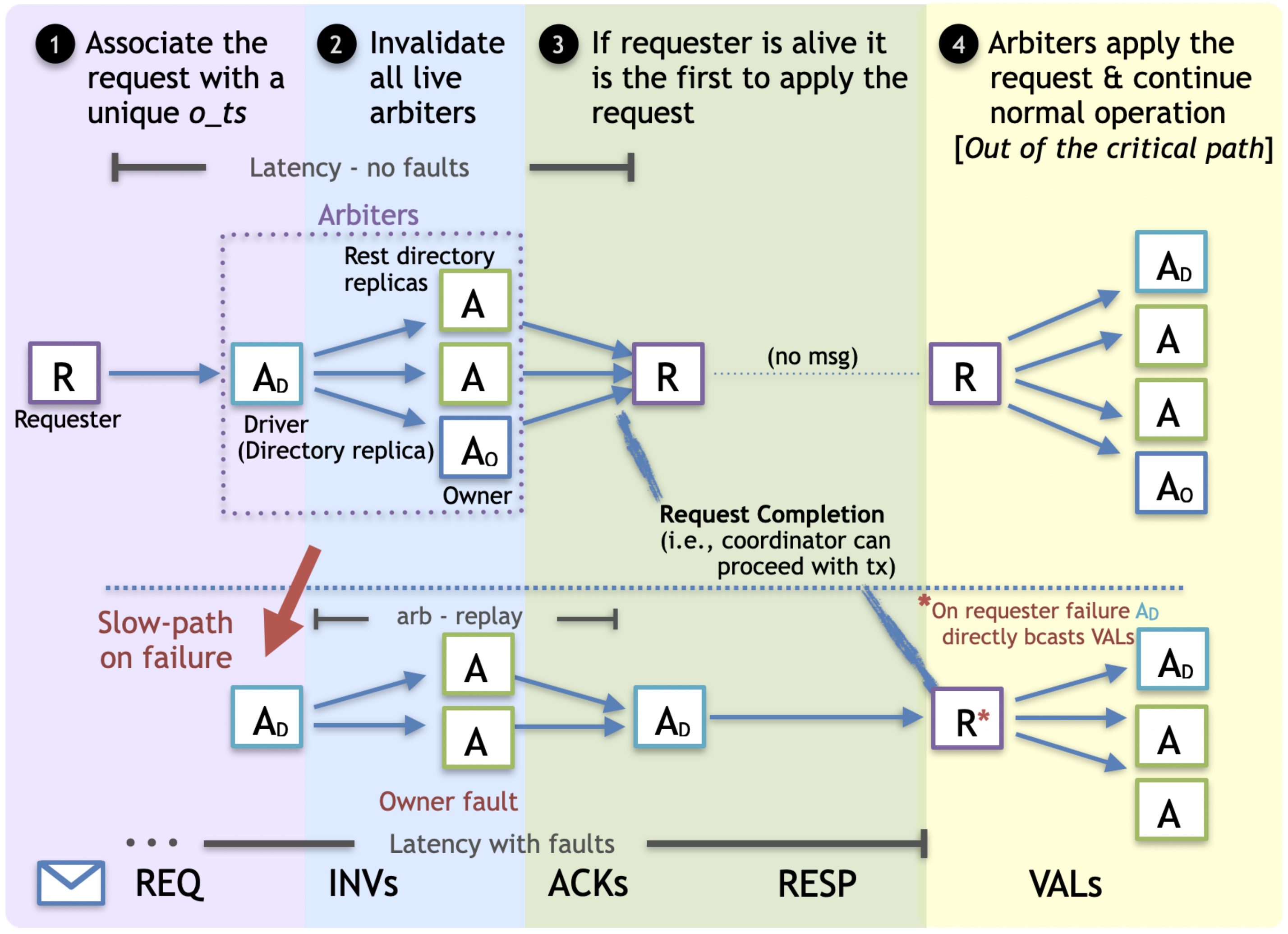}
  \caption{Zeus' ownership protocol with and without faults.}
  \vspace{-10pt}
  \label{fig:sharding-prot}
\end{figure}

\vskip 0pt \noindent{\bf Failure- and contention-free operation.}
An ownership request is illustrated at the top of Figure~\ref{fig:sharding-prot}.
The coordinator that starts a request is called a \textit{requester} node. 
The requester assigns a locally unique request id to the request (to be able to match the response) and sets the object's local $o\_state = Request$.
It then sends a \textit{request} (\CAP{REQ}) message with the request id to an arbitrarily chosen directory node, and this node becomes the {\em driver} of the request. 
The directory nodes and the object owner help arbitrating concurrent ownership requests to the same object, and are called {\em arbiters}.

Upon reception of a \CAP{REQ} message, the driver assigns an ownership timestamp $o\_ts$ to the object and sets its local state to $o\_state = Drive$ \scalebox{0.9}{\circled{1}}. 
It also 
sends an \textit{invalidation} (\CAP{INV}) message containing both the request id and ownership metadata 
to the remaining arbiters (including the current owner)~\scalebox{0.9}{\circled{2}}. 
Assuming no contention for the ownership of the object, each arbiter sets the object's local state to $o\_state = Invalid$, updates its local $o\_ts$ and $o\_replicas$ and responds with an \CAP{ACK} message directly to the requester. 
Note that we optimize the ownership latency by sending the responses directly to the requester instead of passing via the driver. 
If the requester is a non-replica and does not have the data of the object, the current owner includes the data in her \CAP{ACK}.

When the requester receives all expected \CAP{ACK} messages, it applies its request locally before responding to all arbiters with a \textit{validation} (\CAP{VAL}) message~\scalebox{0.9}{\circled{3}}. To apply the request, \newtext{it updates the $o\_replicas$ to specify itself as the new owner,} and sets its object's local $o\_state = Valid$.
Finally, upon reception of the \CAP{VAL} message, each arbiter also applies the request in the same way and the request is finished~\scalebox{0.9}{\circled{4}}.

Notice that to keep $o\_replicas$ consistent with the replica placement and the access levels of the object, the requester must apply the request before any of the arbiters. Moreover, once the requester receives all the \CAP{ACK} messages, it unblocks the application. Thus, the application resumes its transaction after 1.5 round-trips, as shown in the top part of Figure~\ref{fig:sharding-prot}.

\vskip 2pt \noindent{\bf Contention resolution.}
Zeus uses the $o\_ts$ timestamp to resolve contending requests. 
Multiple nodes may issue an ownership request for the same object concurrently through different drivers. 
Each driver creates a per-object unique timestamp for the request $o\_ts =$<$obj\_ver + 1, node\_id$>, using its previous local $obj\_ver$ and own $node\_id$\vspace{1pt}~\scalebox{0.9}{\circled{1}}. 
In case of contention, a driver of one of the contending requests will receive an \CAP{INV} message of another contending request (for the same object) \scalebox{0.9}{\circled{2}}. 
It will only process the \CAP{INV} message if the $o\_ts$ in the message is lexicographically larger than its own $o\_ts$ for the object.
This guarantees that there is one and only one winner of each contention. 
All the drivers whose requests fail send a \CAP{NACK} message to their requesters.
Similarly, the owner responds with a \CAP{NACK} directly to the requester if the requested object is involved in a pending transaction (Section~\ref{sec:zeus-protocol}).
Upon receiving a \CAP{NACK} the requester either aborts its ownership request or retries it later. 

\vskip 2pt \noindent{\bf Failure recovery.}
The failure recovery procedure starts when the reliable membership is updated after fault detection and the expiration of leases. Each live directory node (and the live owners) update their $o\_replicas$ removing any non-live nodes. The objects whose owners died will be taken over by a new owner on the next write transaction. 
After the membership update which increases the epoch id ($e\_id$), 
requests from previous epochs are ignored. 
This is achieved by including the $e\_id$ of the current epoch in the \CAP{INV} and \CAP{ACK} messages. \newtext{The requester and arbiters ignore these type of messages when their $e\_id$s differ from their local ones}.

A node fault followed by a membership update can leave arbiters of a pending ownership request in \textit{Invalid} $o\_state$.
Nevertheless, any arbiter has all the information to replay the \newtext{idempotent} arbitration phase of the ownership request (dubbed \textit{arb-replay}) between the live arbiters and unblock. A blocked arbiter acts as the request driver and initiates an \textit{arb-replay} by constructing and transmitting the same exact \CAP{INV} message using its local state.
During \textit{arb-replays} some arbiter may receive an \CAP{INV} message for a request it has already applied locally (with same $o\_ts$). In this case, the arbiter simply responds with an \CAP{ACK}.
A basic recovery path from an owner failure is illustrated at the bottom of Figure~\ref{fig:sharding-prot}.

Note that in the recovery process the arbitration phase of an ownership request is finalized with \CAP{ACK} messages sent from the arbiters to the driver instead of the requester, as shown in Figure~\ref{fig:sharding-prot}. 
This is done in order to have a single recovery process that covers failures of all nodes including the requester. 
If the requester is not live the driver directly sends \CAP{VAL} messages to unblock the other live arbiters. 
Otherwise, for safety, as in the failure-free case, the requester must be the first to apply the request. To achieve that we introduce a new \CAP{RESP} message which confirms the win of the arbitration to the requester; who can then apply the request prior to sending \CAP{VAL} messages to the live arbiters, as before.

\subsection{Fast scalable ownership}

The Zeus ownership protocol is \textit{scalable} since it 1)~does not store directory metadata for each object at every transactional node; 
2)~does not broadcasts to every transactional node to locate an object's owner.
Zeus' ownership protocol has a latency of at most 3 hops (without faults and contention) to reliably acquire the ownership regardless of the node requesting the ownership.
We believe this to be the lowest possible latency for a \textit{scalable} ownership protocol. 
The worst-case latency is incurred when an ownership request originates from a non-replica node where neither the owner nor the requester are co-located with the object's directory metadata.
To proceed, the requester must receive the latest value of the object. In order to locate the object, the requester should first contact the directory. The directory will forward the request to the owner, which, in turn, will send the value to the requester, resulting in 3 hops. Note that if the requester is co-located with a directory replica, the first hop is eliminated and ownership is acquired after just one round-trip (2 hops) to the owner.
\section{Reliable commit}
\label{sec:zeus-protocol}

\begin{figure}[t]
  \vspace{-5pt}
  \includegraphics[width=0.925\columnwidth]{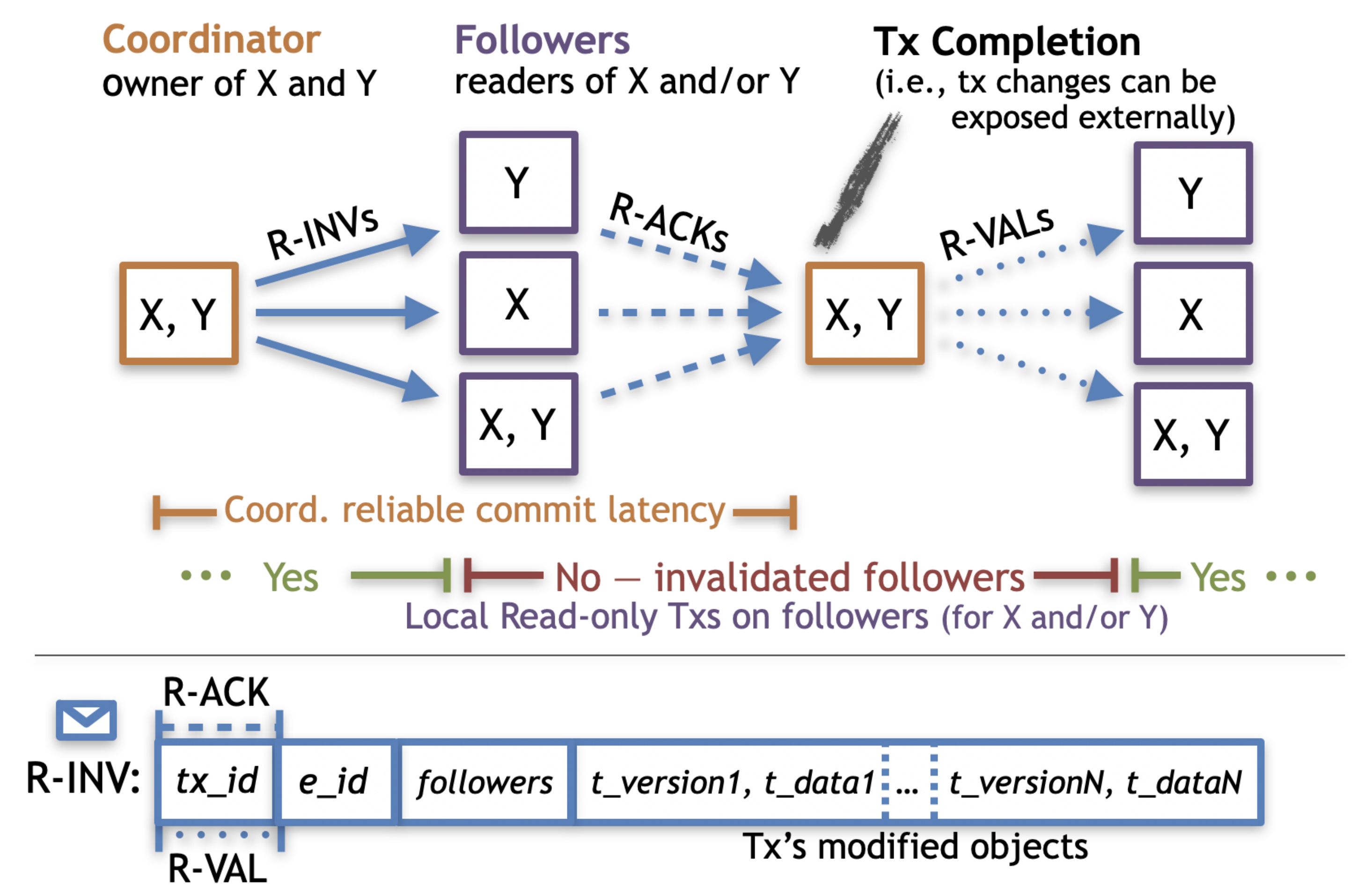}
  \vspace{-11pt}
  \caption{\newtext{Zeus' reliable commit protocol and its messages.}}
  \vspace{-12pt}
  \label{fig:reliable_commit}
\end{figure}

Zeus reliable commit protocol is responsible for propagating the updates made by a local transaction to all of the followers (illustrated in Figure~\ref{fig:reliable_commit}). 
For clarity, we start by describing the information maintained by the protocol. 
We next overview the operation without faults, and then discuss the case with failures. 
Finally, we present two optimizations: pipelining and local read-only transactions from all replicas.

\vskip 4pt \noindent{\bf(Meta)data.}
Each replica (i.e., the owner and readers) keep the following information for an object:
\squishlist
\item $t\_state$: the state of the object, which can be either \textit{Valid}, \textit{Invalid} or \textit{Write};
\item $t\_version$: the version of the object, which is incremented on every transaction that modifies the object;
\item $t\_data$: the data of the object stored by the application.
\squishend

\noindent For every transaction, at the beginning of reliable commit, the coordinator generates a unique $tx\_id =$ <$local\_tx\_id,$ $node\_id$>, where $node\_id$ is its own id and $local\_tx\_id$ is a locally unique, monotonically increasing transaction id.

\subsection{Reliable commit protocol}
\label{sec:reliable_commit_protocol}
\vskip 0pt \noindent{\bf Failure-free operation.} 
At the end of the Local Commit phase, the transaction coordinator updates the $t\_data$ of all modified objects with its private copies created during the Prepare \& Execute phase. 
It also increments their $t\_versions$ and sets $t\_state = Write$ \newtext{--- for pending reliable commit}.

At the beginning of the Reliable Commit phase, the coordinator broadcasts an {\em invalidation} (\CAP{R-INV}) message to all followers.
\newtext{As shown at the bottom of Figure~\ref{fig:reliable_commit},}
this message contains the $tx\_id$, the current epoch id ($e\_id$) and the $node\_id$s of all followers.
For each updated object, it also contains the new $t\_version$ and $t\_data$.
The coordinator temporarily stores the \INV message locally.

Upon receiving an \INV message, a follower checks if the received and the local $e\_id$ match, if not the message is ignored.
\newtext{If they match, the follower goes through each updated object and compares its local $t\_versions$ with that of the message. In case an object's local version is greater or equal, it skips the update of that object. Otherwise, it updates the local $t\_data$ (the actual content of the object) and $t\_version$ with the new ones from the message, and sets its local $t\_state = Invalid$ --- denoting that the object has a pending reliable commit}.
A follower then responds to the coordinator with an \ACK message containing the same $tx\_id$ 
and temporarily stores the \INVNOSPACE.

Once the coordinator receives {\ACKNOSPACE}s from all the followers, 
it reliably commits the transaction locally by changing the $t\_state$ of each updated object to \textit{Valid}. 
Subsequently, the coordinator broadcasts a \textit{validation} (\CAP{R-VAL}) message containing the $tx\_id$ to all followers and discards the previously stored \INV message of the transaction.
When a follower receives an \VAL message for which it has already stored an \INV message (with same $tx\_id$), it sets the $t\_state$ of all objects previously updated by the transaction to the \textit{Valid} state if and only if their $t\_version$ has not been increased. 
It then discards the stored \INV message.

\vskip 2pt \noindent{\bf Reliable replay under failures.}
A node failure triggers a membership reconfiguration, where the epoch id ($e\_id$) is increased and the set of live nodes is updated. 
Subsequently, the ownership protocol stops accepting requests for objects whose owner node is not live in the current membership.

At this point, each locally stored \INV message on any live node represents a pending transaction in the Reliable Commit phase. 
A live node, 
replays its own pending reliable commits and those from the failed nodes. 
This is accomplished by first updating the local pending \INV messages (issued or received) with the new $e\_id$ and 
\newtext{by removing all non-live nodes from followers.}
The messages are then re-sent and handled as explained before. 
A follower who receives an \INV message with the latest $e\_id$ for a transaction ($tx\_id$) that it has previously stored locally simply ignores its content and responds with an \ACK.
Although multiple nodes may replay the reliable commit phase of the same transaction, all relevant \INV messages \snewtext{are idempotent} containing the same $tx\_id$ (and $t\_versions$) so only one can apply updates. 

When a node has no more pending reliable commits (\INV messages) from nodes that are not live, it informs the ownership protocol that it has finished the recovery (Section~\ref{sec:sharding-protocol}). 
Once all live nodes finish the recovery, the ownership protocol starts accepting again all ownership requests as normal.

\begin{figure}[t]
  \centering
  \includegraphics[width=1.04\linewidth]{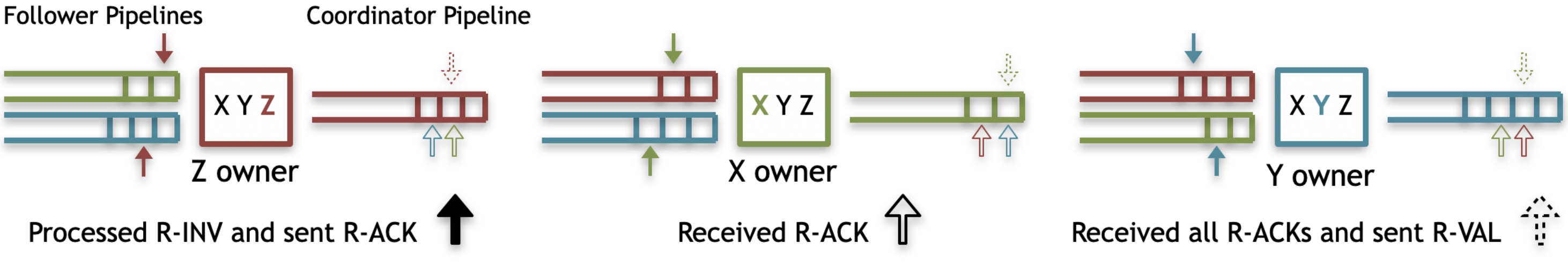}
  \vspace{-20pt}
  \caption{\newtext{Zeus' per-node (in reality per-thread \S\ref{sec:system}) pipelines.}}
  \vspace{-13pt}
  \label{fig:pipe}
\end{figure}    
\subsection{Non-blocking transaction pipelining}
\label{sec:tx-pipelining}
We further introduce transaction pipelining to avoid blocking the application at the coordinator during replication (illustrated in Figure~\ref{fig:zeus_pipe}).
This is possible because a locally (unreliably) committed transaction at the coordinator cannot be aborted. 
Thus, the coordinator can proceed using its locally committed values with certainty. 

However, Zeus also needs to maintain the strict serializability 
on each backup replica. 
Thus, it requires that followers respect the pipeline order of the coordinators when applying updates.
For this, Zeus uses $tx\_id =$<$local\_tx\_id, node\_id$> which is transmitted in every \INV message and contains both the local transaction order within the node $local\_tx\_id$ and the $node\_id$. 
As a result, although there could be several pending causally-related reliable commits, all will be applied in the correct order as specified by the $local\_tx\_id$.

Note that the ordering is enforced only within each different pipeline \newtext{as shown in Figure~\ref{fig:pipe}}.
This is because an object's owner change (i.e., when an object switches pipelines) is not approved until all pending reliable commits with that object have been completed (Section~\ref{sec:sharding-protocol}). 
Thus, an object cannot be involved in pending transactions from two different coordinator nodes and the ordering across coordinators does not matter. 
We further optimize this by enabling per-thread (instead of per-node) pipelines via our choice of local commit as explained in Section~\ref{sec:system}.
The pipelining optimization also reduces the number of \ACK and \VAL messages, since sending a message with a $tx\_id$ implies the successful reception and processing of all previous messages in that pipeline.

\begin{comment}
For correct serialization after failures, each coordinator also keeps track of followers that store any of its pending \INV messages, and forwards all new \INV messages to them as well. This enables followers to replay all stored \INV messages from failed coordinators in the order they were originally issued and recover safely. 
\end{comment}

% \textcolor{red}{
A node may not be a follower of all \CAP{R-INV}s, and thus may receive only a partial stream of a pipeline.
An extra condition is needed for when such followers can \textit{apply} an \CAP{R-INV}.
A follower applies an \CAP{R-INV} if for the previous $local\_tx\_id$ (slot) of the pipeline it has either applied an \CAP{R-INV} or has received an \CAP{R-VAL}. The latter occurs 
for a transaction follower \textit{F} who was not also a follower of the previous slot in the pipeline. 
To facilitate this, during the broadcast of an \CAP{R-INV}, the coordinator piggybacks a \textit{prev-\CAP{VAL}} bit if it has broadcasted \CAP{R-VAL}s for the previous slot.
Otherwise, it includes \textit{F} in the \CAP{R-VAL} broadcast of that previous slot.
Finally, after a coordinator's failure, an \CAP{R-INV} is considered as a pending reliable commit and is replayed by a follower {\em iff} that follower has not only received but also applied the \CAP{R-INV} message. 
% }

\begin{figure}[t]
  \centering
  \includegraphics[width=0.95\linewidth]{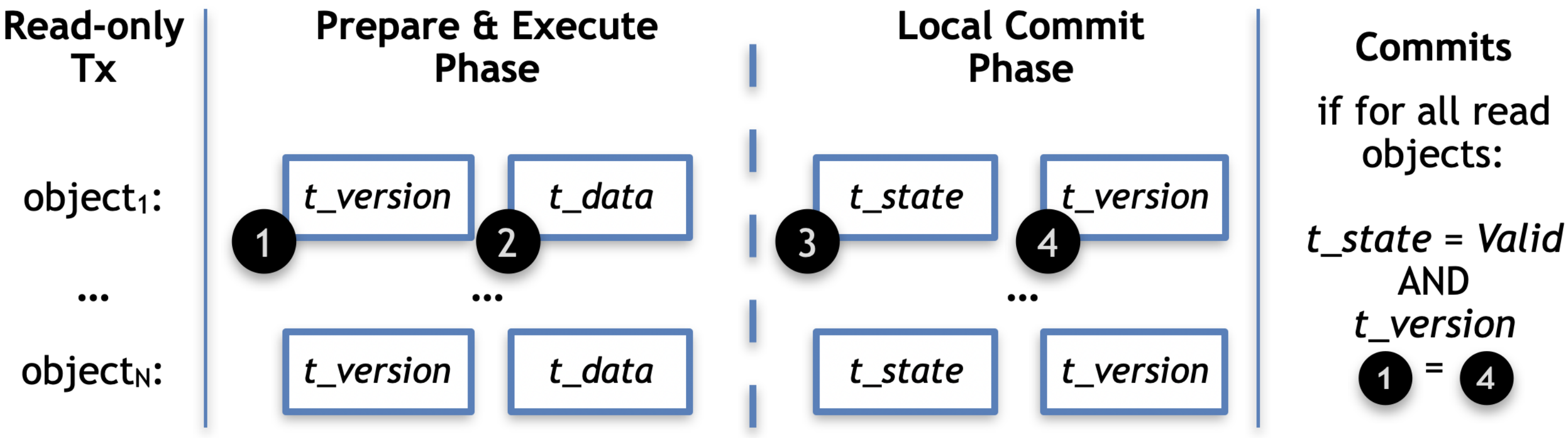}
  \vspace{-10pt}
  \caption{\newtext{Zeus' consistent read-only transactions on readers.}}
  \vspace{-13pt}
  \label{fig:read-only}
\end{figure}
\subsection{Read-only transactions}
Zeus optimizes read-only transactions by allowing them to be executed locally from any replica that stores all relevant objects, regardless of the ownership level (read or write), and without compromising strict serializability. 
\newtext{
This is enabled by three ideas.
First, read-only transactions do not need to communicate any updates to other replicas. 
Second, a verification-based scheme can be applied to exploit the local object versioning
and ensure a consistent snapshot across all reads of a read-only transaction.
Finally, the reliable commit guarantees that all replicas are invalidated before any updated state is exposed externally by the readers. 
We elaborate on the latter before discussing the read-only protocol.}

\vskip 3pt \noindent{\bf Invalidation-based reliable commit.}
A locally committed write transaction does not reliably commit on the owner unless it has invalidated all its followers (i.e., the readers of modified objects).
As noted before, a reader which applies an invalidation to its local object also updates its object's local value with the newly received value. Thus, it cannot return neither the old value nor the new one as the object has been invalidated.
The reader can return the new value only after it receives the \CAP{R-VAL} message and validates its local object.

Simply put, there is a transitioning period until a reader can safely return the new value.
That period ends once all readers of a modified object have stopped returning the old value and have received the new one.
If a reader was to prematurely return the new value (i.e., before receiving the \CAP{R-VAL} message and the end of that period), two things could go wrong. First, another reader who has not yet invalidated the object could subsequently return the old value and compromise consistency. 
Second, if all nodes that have received the new (not yet reliably committed) value fail\footnote{That is a smaller number of nodes than the replication degree.}, then the prematurely returned value would be permanently lost. 

\vskip 1pt \noindent{\bf Read-only protocol.}
Consequently, in Zeus, a read-only transaction completes after only two phases as shown in Figure~\ref{fig:read-only} and described next.
In the Prepare \& Execute phase, the coordinator of a read-only transaction sequentially reads and buffers the $t\_version$ and the value ($t\_data$) of each local object as specified by the transaction.
In the Local Commit phase, the coordinator checks if all accessed objects are in $t\_state = Valid$ before verifying that all $t\_versions$ have remained the same.
If so, the transaction commits successfully. 
Otherwise, there is an ongoing conflicting (local or remote) reliable commit and the read-only transaction is aborted.

\vskip 1pt \noindent{\bf Use-case.}
\newtext{Apart from the obvious performance benefit, one example where the read-only optimization is useful is con\-trol/data-plane applications, such as in a cellular networks. There, write transactions are executed by a control-plane node (the Zeus owner), for instance to configure routing, while all data-plane nodes (i.e., Zeus readers) can perform consistent read-only transactions locally,  e.g., for forwarding.}

% \vspace{-5pt}
% \vspace{-10pt}
\section{Discussion}
\label{sec:discussion}
\subsection{Distributed commit vs Zeus}
\newtext{
Traditional datastores statically shard objects and execute reliable transactions in a distributed manner across servers.
% for scalability and reliability. 
This poses two challenges. The first 
is accessing the objects. 
Static sharding schemes do not guarantee that all objects accessed by a 
transaction reside on the same node. Frequently, one or more objects 
in a transaction are stored remotely. In this case the execution stalls until the objects are fetched -- sometimes sequentially (e.g, pointer chasing or control flow).} 

\newtext{
The second challenge is handling concurrent transactions on conflicting objects. If two nodes try to commit transactions on conflicting objects simultaneously, one of them has to abort. Detecting and handling these conflicts under the uncertainty of faults needs extra signaling across nodes. Thus, transactional systems 
based on distributed commit 
need numerous round-trips to commit each transaction (e.g., see FaSST). Moreover, a node cannot start the next transaction on the same set of objects until the commit is finished, as it cannot be sure that it will not have to abort. This introduces several round-trips of delay in the critical path of the commit and significantly reduces the transactional throughput.}

\newtext{
Zeus replaces remote accesses and distributed commit with its (occasional) ownership, local accesses and reliable commit to addresses the two main issues mentioned above and accelerate workloads with locality.
Firstly, the ownership makes objects accessed by a transaction accessible locally most of the time, which avoids stalls during the execution. 
Secondly, only a single node (the owner) can execute a write transaction on an object at a time, so a transaction cannot be aborted remotely, commits after a single round-trip and is pipelined. 
Zeus' reliable commit also allows local and consistent read-only transactions from all backups.}

\newtext{
Unlike distributed commit, Zeus' ownership is a protocol specialized for single-object atomic operations (including migration).
Zeus resolves concurrent ownership requests in a decentralized way, and applies an idempotent scheme to tolerate faults without extra overheads on the common failure-free case. This makes acquiring ownership reliable yet fast (1.5 round-trips) during fault-free operation.}

\vspace{-8pt}
\subsection{Other details}
\vspace{-2pt}
\beginbsec{Cost of ownership vs. remote access}
\newtext{
The object size influences the cost of acquiring ownership for it by a non-replica node similarly to a remote access, since in the fault-free case the value is included in a single ownership message as in the response of a remote access.
A reader acquires the ownership without the value and thus is not influenced by its size.
The reliability of Zeus' ownership comes with a higher message cost compared to a remote access. These are small constant messages with cost amortized over several local accesses in workloads with locality. 
Nevertheless, for workloads without enough locality, that cost renders Zeus less suitable than remote accesses and distributed commit.}

% \vspace{3pt}
\beginbsec{Deadlocks}
\newtext{
Zeus currently circumvents deadlocks via a simple back-off mechanism. 
For Zeus, such a situation may arise only early in a transaction (i.e., in the Prepare \& Execute phase)
-- when requesting ownership for an object. This manifests with repeated failed ownership requests, after which Zeus aborts and retries a transaction with an exponential back-off. In practice, deadlocks in Zeus are rare because transactions on the same object are mostly executed on the same server by virtue of load balancing.
For deployments where that is not the case, a more sophisticated scheme such as the one proposed by Lin \textit{et al.}~\cite{lin2016towards} may be considered.
}

% \vspace{3pt}
\beginbsec{Distributed directory}
\newtext{
For simplicity, Zeus 
uses a single directory 
for all objects in the deployment. 
The directory is replicated for fault-tolerance, and the ownership protocol is lightweight and is designed to balance the load across all the directory replicas.
However, a single replicated directory may become a scalability bottleneck at large deployment sizes or when locality is limited. In such cases, a distributed directory scheme (i.e., using consistent hashing on an object to determine its directory nodes) should be used instead.}

% \vspace{3pt}
\beginbsec{Sharding request types}
\newtext{
Zeus exploits the ownership protocol for other types of sharding requests, such as reliably removing a reader. For example, when a non-replica acquires the ownership of an object, the total number of replicas increases. 
To keep the initial replication degree and avoid increasing the cost of reliable commits,
we invoke the ownership protocol out-of-the-critical-path to discard a reader.}
% (an object replica).}

% \begin{comment}
% \vspace{3pt}
\beginbsec{Write transactions with opacity}
\newtext{
Apart from strict serializability, Zeus provides an additional guarantee that all write transactions see a consistent snapshot of the database even if they abort. This is also referred to as \textit{opacity}~\cite{Guerraoui:08}. Opacity further enhances Zeus' programmability since by preventing inconsistent accesses in write transactions it relieves the programmer from the effort of handling those cases.}
% \end{comment}
\vspace{-5pt}
\section{System}
\label{sec:system}

We have built a custom in-memory datastore and implemented the Zeus protocols on top of it. 
In this section we briefly discuss the implementation details. 

An application communicates with the datastore through a transactional memory \CAP{API} that consists of primitives to create and manage memory objects of different sizes.
This includes implementations of \mylisting{malloc} (create an object), \mylisting{free} (destroy an object), 
\mylisting{tr\_open\_read} and \mylisting{tr\_open\_write} (for marking object as used in a transaction for reading and writing). 
Each transaction starts with a create transaction call \mylisting{tr\_create} (for write) or \mylisting{tr\_r\_create} (for read-only transaction), followed by an arbitrary code that can invoke the above \CAP{API}s, and finishes with a \mylisting{tr\_commit} (or \mylisting{tr\_abort}), at which point the local commit starts (aborts). 
This is a low-level \CAP{API}, very similar to the one used by FaRM, and it allows great flexibility to build further abstractions on top of it.

The datastore is implemented in C over \CAP{DPDK}, and it consists of two parts. 
One part is the datastore module that runs as a separate process implementing the main datastore functionality.
The other part is the Zeus library that is linked to any application over shared memory without limiting its architecture (it can be a separate process, container, etc).

The datastore module implements the Zeus protocols, the transactional memory \CAP{API}, and a reliable messaging between nodes. 
Zeus communicates between nodes using a custom reliable messaging library we built on top of \CAP{DPDK}.
The datastore module also includes a customizable, application-aware load balancing functionality, as described in Section~\ref{sec:design}.

Both application and the datastore modules can run in multiple threads. 
In the evaluation, we use up to 10 application and 10 datastore worker threads.
These threads are pinned to their own cores. We also use one core for \CAP{DPDK}.

We implement a simple multi-threaded Local Commit (Section~\ref{sec:design}) using the same intuition as for the overall Zeus. 
Each thread that executes a transaction has to become the owner of each object. 
However, this ownership is local and managed through standard locking. 
We leverage the aforementioned load balancer to enforce locality across the threads and increase concurrency.  
Apart from simplicity, this also enables transaction pipelining to be applied on a per-thread bases which increases the overall concurrency of reliable commits.

Currently, porting an application to Zeus requires manual code modification on pointer accesses, similarly to prior work (e.g., as in FaRM). 
However, this can be automatized at a compiler level, as performed by Sherry \textit{et al.}~\cite{Sherry:15}. 
\section{Evaluation} 
\label{sec:evaluation}

\begin{table}[t!]

\resizebox{\columnwidth}{!}{% use resizebox with textwidth
\begin{tabular}{l|lrrrr}
 &
  \textbf{characteristic} &
  \multicolumn{1}{l}{\textbf{tables}} &
  \multicolumn{1}{l}{\textbf{columns}} &
  \multicolumn{1}{l}{\textbf{txs}} &
  \multicolumn{1}{l}{\textbf{read txs}} \\ \cline{2-6} 
\textbf{Handovers} & large contexts  & 5 & 36 & 4 & 0\%  \\
\textbf{Smallbank} & write-intensive & 3 & 6  & 6 & 15\% \\
\textbf{TATP}      & read-intensive  & 4 & 51 & 7 & 80\% \\
\textbf{Voters}    & popularity skew & 3 & 9  & 1 & 0\% 
\end{tabular}
}

\vspace{5pt}
\caption{Summary of evaluated benchmarks.}
\vspace{-20pt}
\label{tab:bench}
\end{table}

\beginbsec{Formal verification}
We specified the ownership protocol and the reliable commit of Zeus in \CAP{TLA$^{+}$} and model checked them in the presence of crash-stop failures, message reorderings and duplication. 
We have verified them against several key invariants including the following:
\squishlist
\item Live nodes\footnote{By construction non-live nodes cannot compromise safety because $e\_ids$ prevent them from participating in either transaction or ownership requests.} in $t\_state$=$Valid$ have always consistent data.
\item All live arbiters in $o\_state$=$Valid$ agree and correctly reflect the owner and reader nodes of the object.
\item At any time there is at most one owner and that owner stores the most up-to-date value of the object.
\squishend 
The detailed protocol specifications and the complete list of the model-checked invariants can be found online\footnote{\href{https://zeus-protocol.com/}{https://zeus-protocol.com}}. 

\begin{figure*}[ht]
\setkeys{Gin}{width=1.05\linewidth}

\begin{tabularx}{\linewidth}{XXX}
    \includegraphics{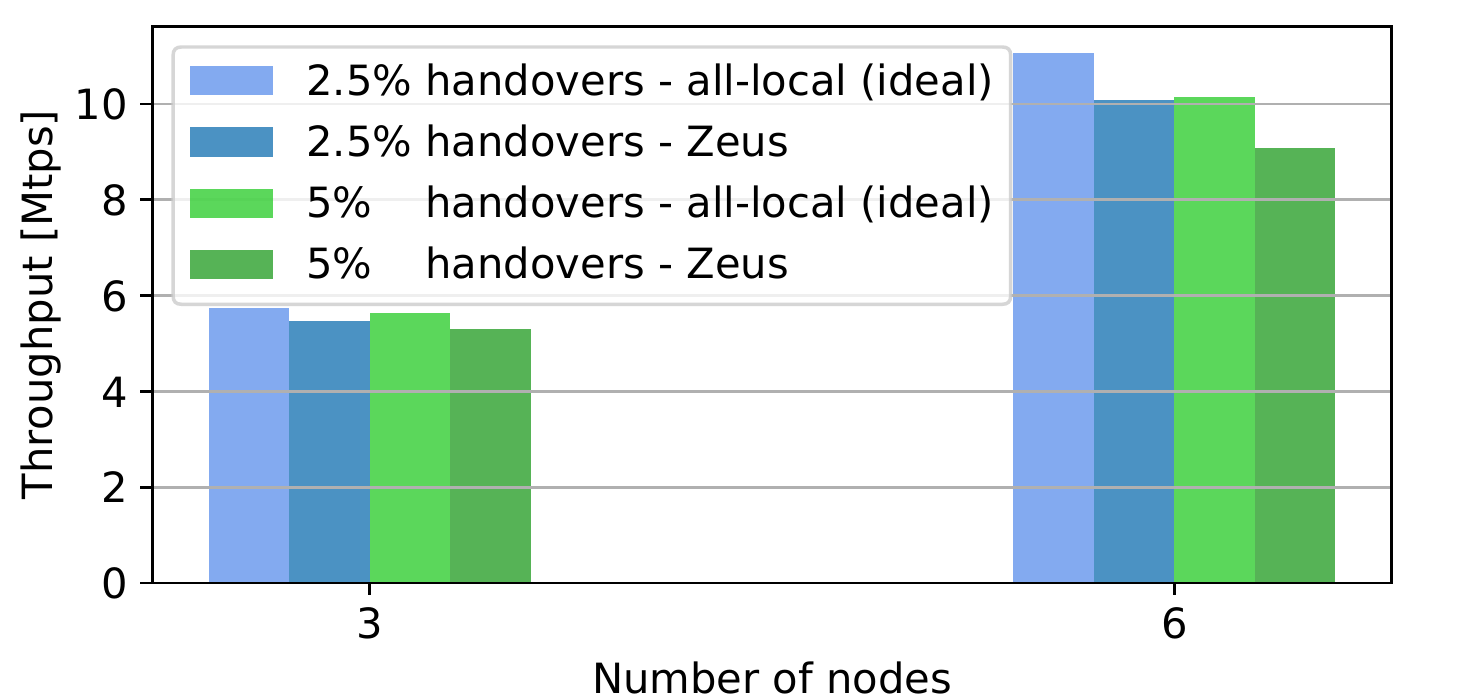}
    \vspace{-18pt}
    \caption{All-local (ideal) vs. Zeus for 2.5\% and 5\% handovers on 3 and 6 nodes.}
    \label{fig:ho_total}
&
    \includegraphics{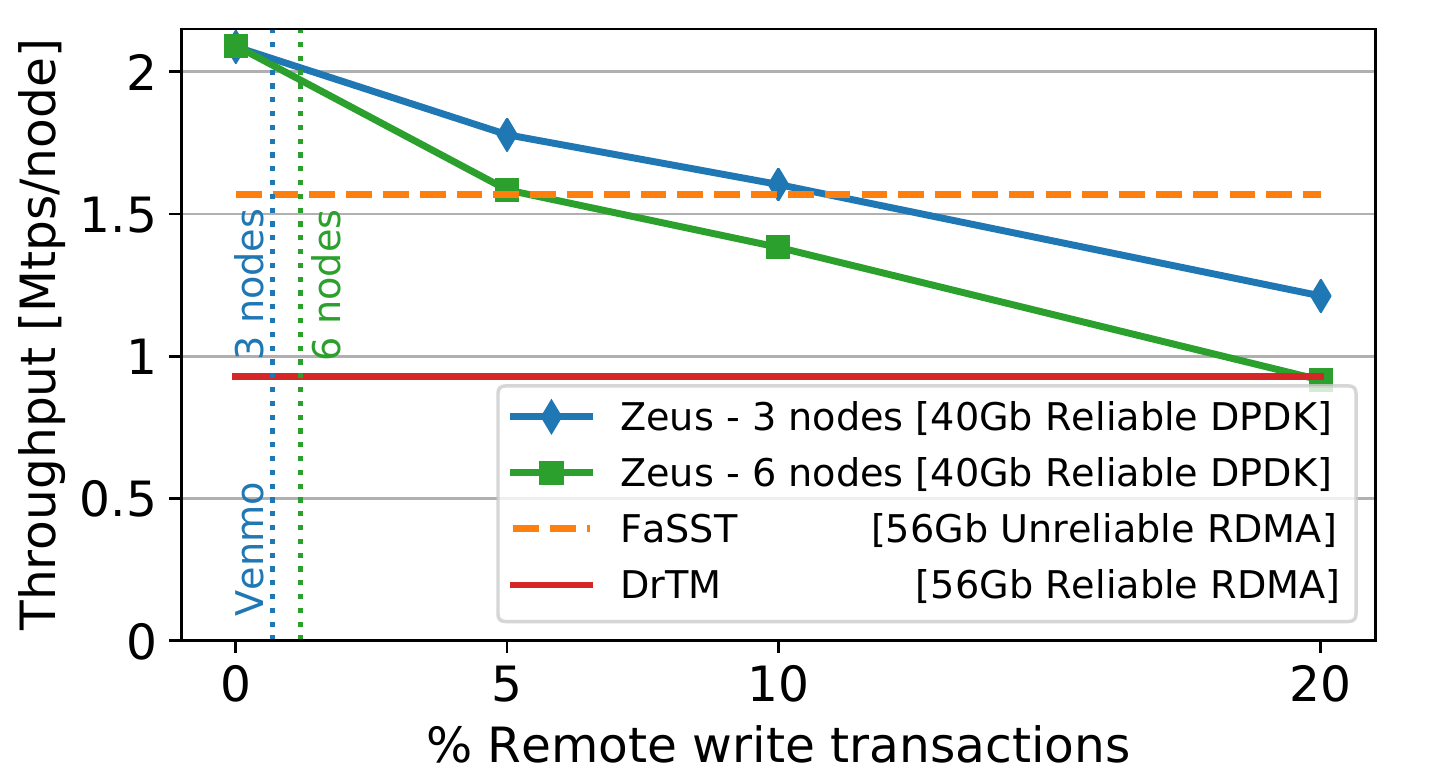}
    \vspace{-18pt}
    \caption{Smallbank while varying remote write transactions. 
    }
    \label{fig:eval_smallbank}
&
    \includegraphics{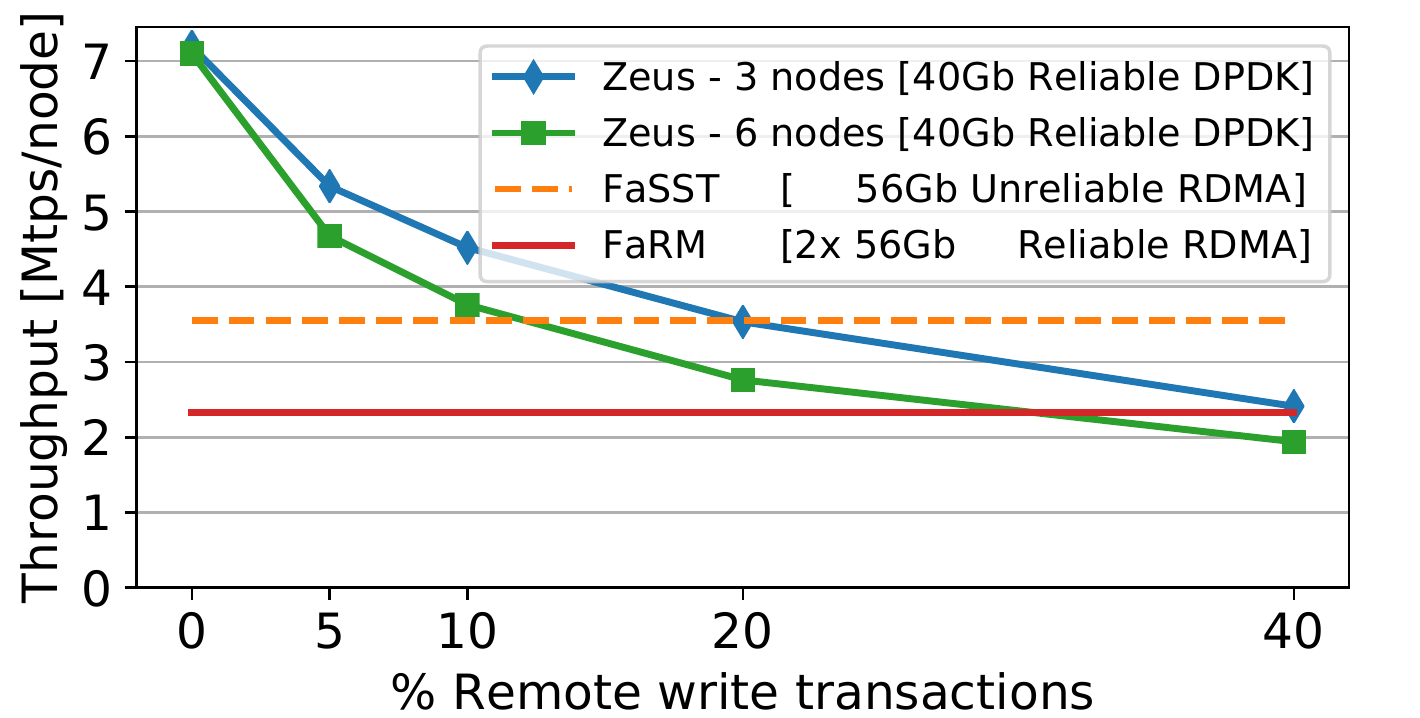}
    \vspace{-18pt}
    \caption{\CAP{TATP} while varying remote write transactions.
    }
    \label{fig:tatp_local}
\end{tabularx}

\vspace{-20pt}
\end{figure*}

% \vskip 3pt \beginbsec{Locality in workloads}
\vskip 2pt \beginbsec{Locality in workloads}
\newtext{We begin by briefly analysing the locality of access patterns in workloads. For this, 
we report the fraction of remote transactions of three workloads, spanning the telecommunication, financial, and trade sectors. 

\squishlist
% \item \vspace{4pt} \textbf{Boston cellular handovers}:
\item \textbf{Boston cellular handovers}:
As explained in Section~\ref{sec:background},
in a cellular workload, remote transactions are caused by remote handovers.
To evaluate the real-world frequency of remote handovers, we use the population and mobility model from Boston metropolitan area~\cite{bostom2013mobility} with the reported averaged daily commute of 100km.
We assume base stations are uniformly spread through the area at a distance of 1km (with a typical coverage of a macro cell~\cite{cell-coverage} and a common ratio of cells per population~\cite{mohammadkhan2016considerations}).
These are sharded across all nodes in a deployment. 
As the number of nodes increases, the number of remote handovers also increases, up to 6.2\% for six nodes.
In summary, for a setup where 5\% of all transactions are handovers
and out of these 6.2\% of handovers are remote (in a six node deployment), there are in total 0.31\% remote transactions. 

% \item \vspace{4pt} \textbf{Venmo transactions}:
\item \textbf{Venmo transactions}:
We use the most recent public Venmo dataset~\cite{venmo-dataset} with more than seven million financial transactions to analyze the fraction of remote transactions. 
We partition the users to nodes, but 
still observe 0.7\% and 1.2\% of remote transactions for 3 and 6 nodes, respectively.

% \item \vspace{4pt} \textbf{TPC-C}:
\item \textbf{TPC-C}:
We mathematically analyze the number of remote transactions in the \CAP{TPC-C} benchmark, which is considered representative for industries that trade products.
In \CAP{TPC-C}, only a small fraction of new-order and payment transactions may result in remote accesses. We find that just 2.45\% of the transactions in the benchmark are remote.
\squishend
We empirically evaluate benchmarks related to cellular and financial transactions (i.e., Handovers and Smallbank). While promising in terms of locality, we leave the experimental evaluation of \CAP{TPC-C} for future work because our current implementation of Zeus does not support range queries.}

\vskip 2pt \beginbsec{Experimental testbed}
We run all our experiments on a dedicated cluster with six servers.
Each server has a dual socket Intel Xeon Skylake 8168 with 24 cores per socket, running at 2.7GHz, 192 GB of \CAP{DDR4} memory and a Mellanox \CAP{CX-3} card. 
We use and pin all our threads into the first socket only, where the network card resides.
All servers communicate through a Dell \CAP{S6100-ON} switch with 40 Gbps links.

We first evaluate Zeus on several benchmarks \newtext{(summarized in Table~\ref{tab:bench}); 
including three benchmarks discussed in Section~\ref{sec:background}
and the \CAP{TATP} benchmark~\cite{TATP:2009} to further study Zeus' limits over FaSST and FaRM.}
\newtext{For benchmarks, as in prior work~\cite{kalia2016fasst}, we consider 3-way replication and enough co-located clients to saturate each evaluated system. The initial sharding of all systems is the same. Unlike Zeus, baselines do not support dynamic sharding (i.e., ownership).}
We were not able to run the baseline systems FaRM, FaSST and DrTM on our platform,
but the hardware used in their evaluation is similar, so we report numbers from their papers~\cite{kalia2016fasst, dragojevic2014farm, wei2015fast}. 
We conclude by demonstrating the ease of porting legacy applications onto Zeus by porting and evaluating a cellular packet gateway, an Nginx server and the \CAP{SCTP} protocol.

\subsection{Handovers}

We start our evaluation with a cellular handovers benchmark. 
We evaluate three operations described in Section~\ref{sec:background}: a handover (consists of two transactions, one at the start and one at the end), a service request and a release (each a single transaction). 
We implement them as defined in \CAP{3GPP} specification, 
on top of Zeus.
All transactions are write transactions. 
A typical cellular phone context for these operation is large and many parts of it get modified so we need to commit about 400B of data per transaction. 

Recall that mobile users perform both handovers and all other requests, while the stationary users only perform other requests (i.e., no handovers). 
In our evaluation, we vary the ratios of the total number of handovers versus the total number of requests (handovers, service requests and releases), each modeling different mobility speeds in the network. 
A typical cellular network has 2.5\% handovers~\cite{mohammadkhan2016considerations}, and we also evaluate the 5\% case corresponding to doubling the mobility. 

We run a benchmark on a population of 2M users out of which 400k are mobile. 
We use the typical cell network provisioning as reported in~\cite{mohammadkhan2016considerations, cell_params}, scaled to 2M users (requiring 1000 base stations). 
Not all handovers will involve ownership transfers because some will occur between objects of the same node.
\newtext{For the ratio of remote handovers we use the numbers we analyzed from the Boston metropolitan area.}

In our evaluation we vary the number of nodes in the system, and plot the total throughput for the two ratios as well as for all local transactions. 
This is shown in Figure~\ref{fig:ho_total}. 
We see that the difference between Zeus and the perfect sharding is at most 9\%. 
This is because there a large fraction of the transactions is local, and we have less than 0.5\% ownership requests.
We also see that the performance scales linearly with the number of nodes, even though there are more transactions with ownership transfers for a larger number of nodes. Lastly, we note that prior works have not studied the handover benchmark; as such, there are no published numbers for state-of-the-art systems to compare against.

\begin{figure*}[t]
\setkeys{Gin}{width=1.00\linewidth}

\begin{tabularx}{\linewidth}{XXX}

    \includegraphics{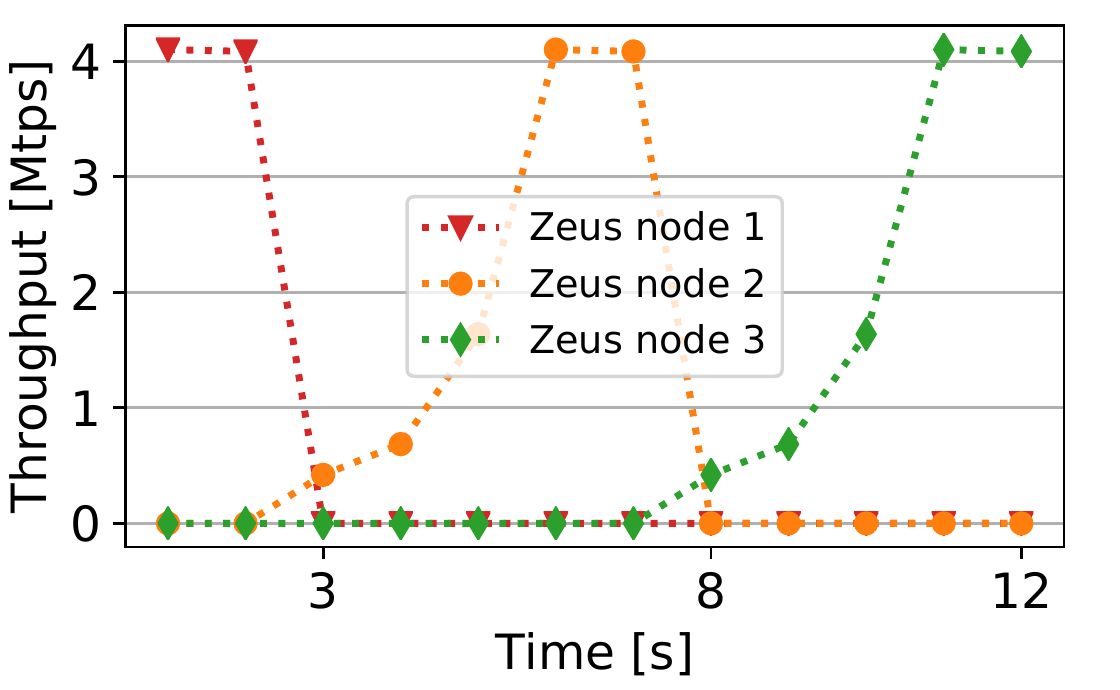}
    \vspace{-18pt}
    \caption{Voter Performance when moving 1M objects across nodes.}
    \label{fig:voter_mov_1}
&
    \includegraphics{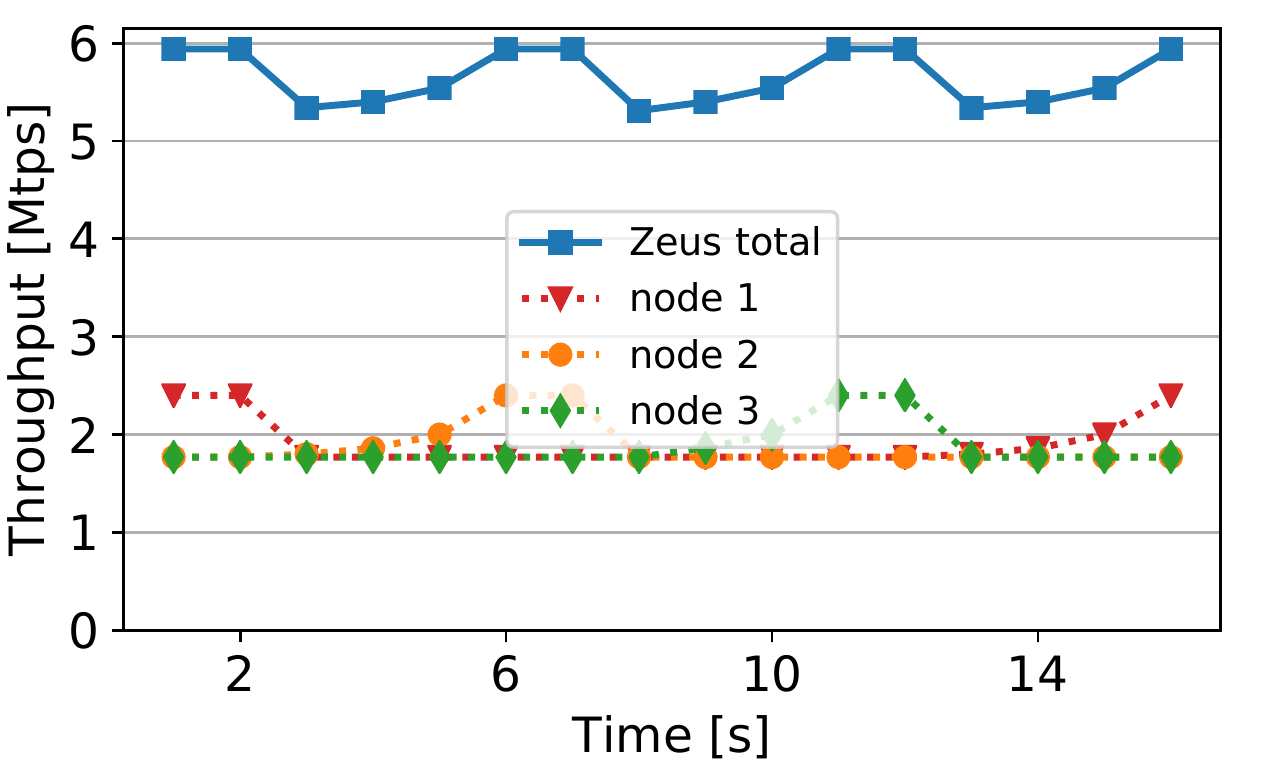}
    \vspace{-18pt}
    \caption{Voter Performance when registering votes and moving objects.}
    \label{fig:voter_mov_2}
&
    \includegraphics{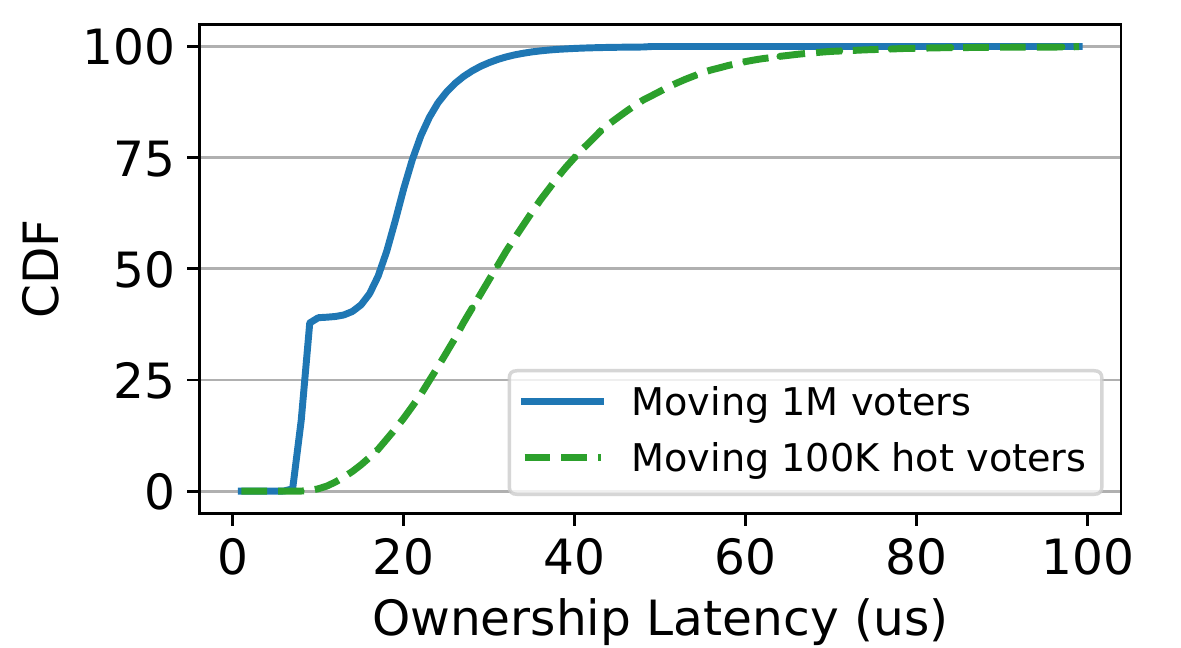}
    \vspace{-18pt}
    \caption{CDF of Zeus ownership request latency for Voter experiments.}
    \label{fig:cdf}
\end{tabularx}

\vspace{-20pt}
\end{figure*}

\subsection{Smallbank}
Smallbank is a benchmark that simulates financial transactions~\cite{cahill2009serializable}.
It is write intensive with 85\% write transactions. 
Out of them, 30\% modify two objects and the rest modify 3 or more objects per transaction.
All read transactions access 3 objects. 
We use the same access skew on objects as in \newtext{FaSST}.

\newtext{Smallbank does not specify which pairs of users transact with each other, hence it cannot be used to infer real-world transaction locality.}
\newtext{
To understand how much the degree of locality affects Zeus, we start increasing the number of transactions that require an ownership change, until Zeus breaks even with the beaselines. 
This is shown in Figure~\ref{fig:eval_smallbank}. 
We see that running Smallbank with the real-world remote transactions, as observed in the Venmo, Zeus outperforms FaSST and DrTM by about 35\% and 100\%, respectively.} 
Recall that neither FaSST nor DrTM support dynamic sharding so any small and gradual change in access pattern will eventually lead to \newtext{an almost random placement and most} requests being remote, which is what we show here.
As expected, Zeus throughput drops as the remote transactions increase and the trend between three and six nodes remains the same. As long as less than 5\% (20\%) of transactions require ownership change, Zeus provides a benefit over FaSST (DrTM).

\vskip 3pt \noindent {\bf Reliable lower-end networking.}
Note that unlike FaSST, Zeus implements reliable messaging with its overheads. 
While this reduces Zeus' performance, it allows Zeus to gracefully tolerate message losses. In contrast, FaSST must kill and recover a node for each lost message.  Also, FaSST uses 56Gb \CAP{RDMA}. DrTM similarly leverages 56Gb \CAP{RDMA} and relies on hardware transactional primitives for its performance. Zeus uses a 40Gb non-\CAP{RDMA} networking and it does not depend on hardware-assisted transactions for its performance.

\vspace{-3pt}
\subsection{TATP}
\vspace{-2pt}

We next evaluate the \CAP{TATP} benchmark \cite{TATP:2009}, which gives us a second point of comparison with other state-of-art systems~\cite{dragojevic2015no, kalia2016fasst}. 
It is read intensive, with 80\% read and 20\% write transactions. 
We use 1M subscribers per server, as in \snewtext{FaSST}.
Similarly to the Smallbank benchmark, we vary the fraction of transactions that require an ownership change. 
The total throughput is shown in Figure~\ref{fig:tatp_local}. 
We see that when the fraction of remote requests is small, Zeus outperforms FaSST and FaRM by up to 2$\times$ and 3.5$\times$ respectively. 

As discussed in the Smallbank study, neither FaRM nor FaSST allow dynamic sharding so they end up issuing remote requests whenever there is a changing access pattern.  
Zeus keeps the requests local by moving objects, and it is especially effective for a read-dominant benchmark like \CAP{TATP}, since there is little overhead on reads. 
We also see that as long as there are fewer than 20\% (40\%) of write transactions with ownership requests, Zeus outperforms FaSST (FaRM).
Again, these thresholds are higher than in the case of Smallbank due to read-dominant workload.
The performance trend of Zeus for three and six nodes is the same as in Smallbank.

\begin{figure*}[ht]
\setkeys{Gin}{width=1.00\linewidth}

\begin{tabularx}{\linewidth}{XXX}

    \includegraphics{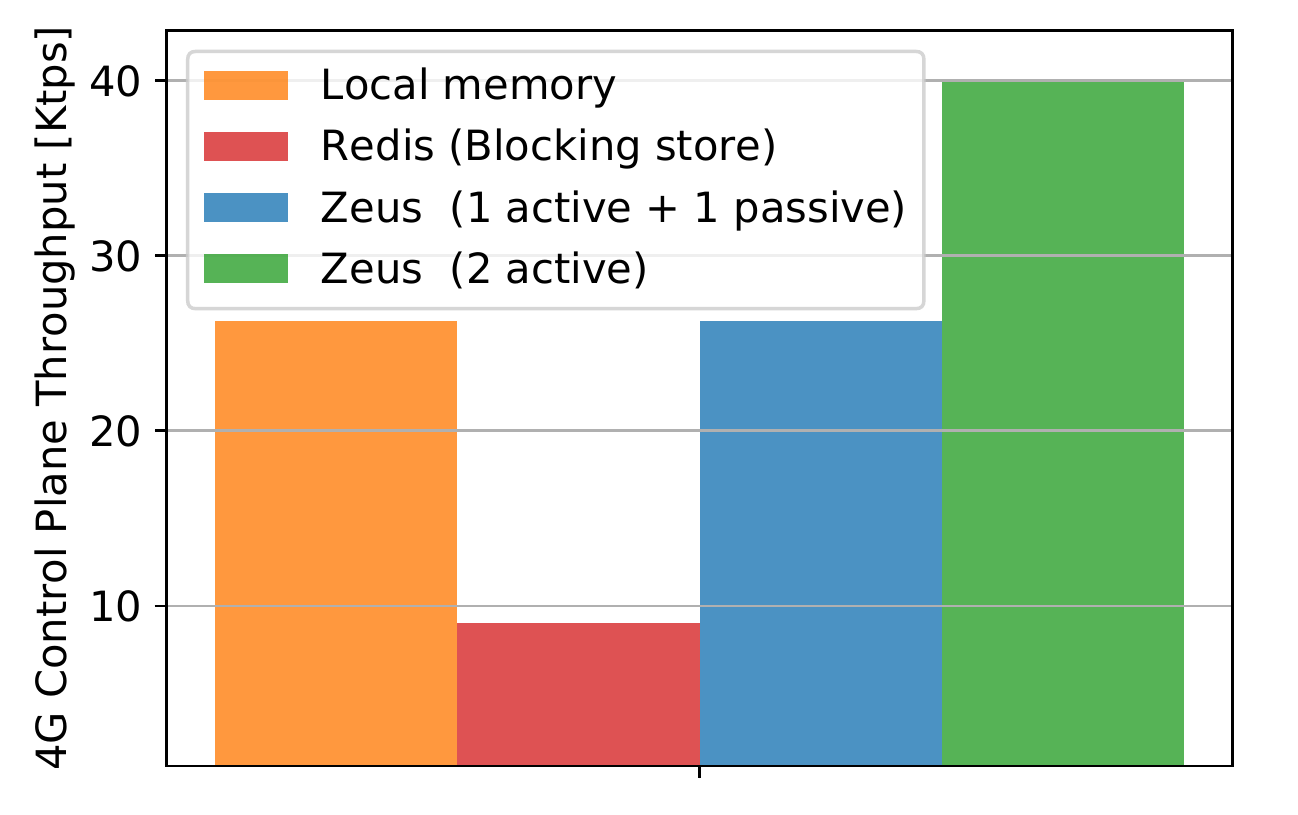}
    \vspace{-18pt}
    \caption{Cellular packet gateway control plane performance.} 
    \label{fig:spgw_thr}
&
    \includegraphics{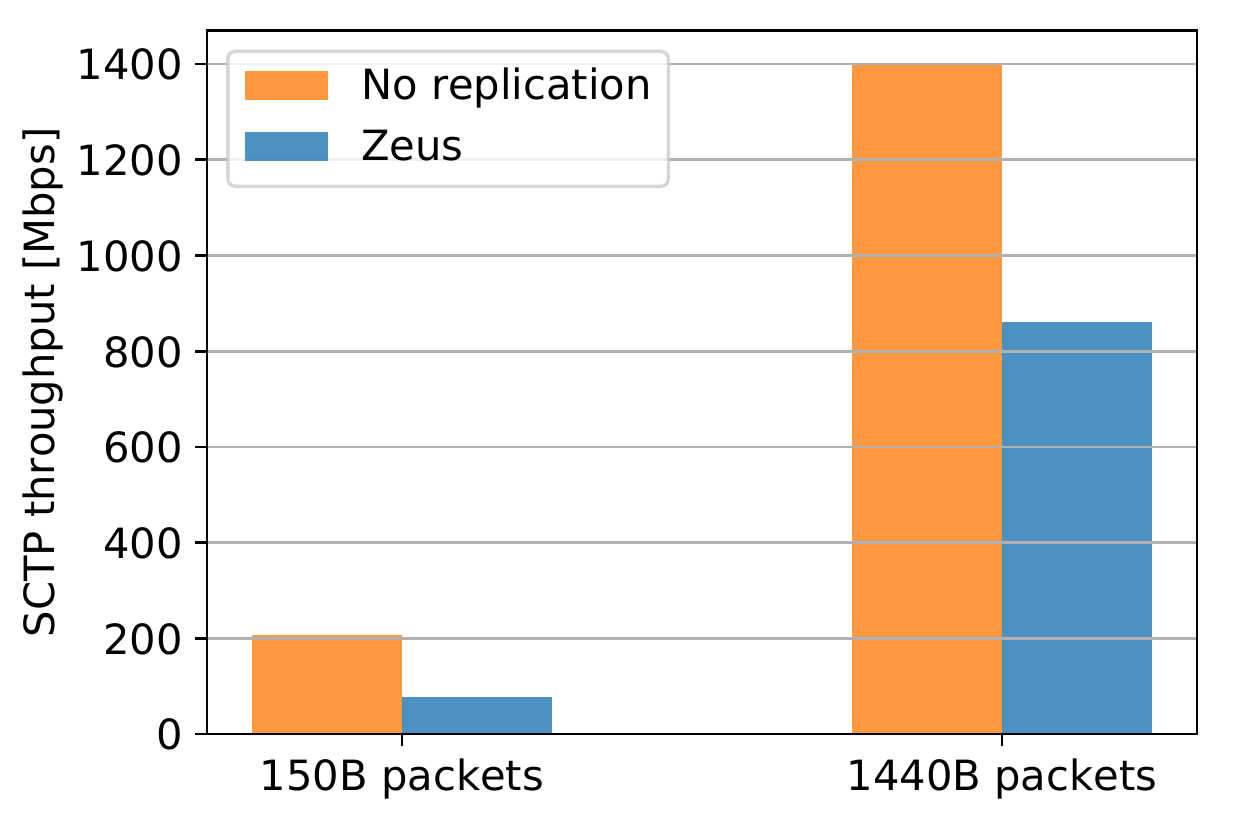}
    \vspace{-18pt}
    \caption{SCTP performance.}
    \label{fig:sctp_thr}
&
    \includegraphics{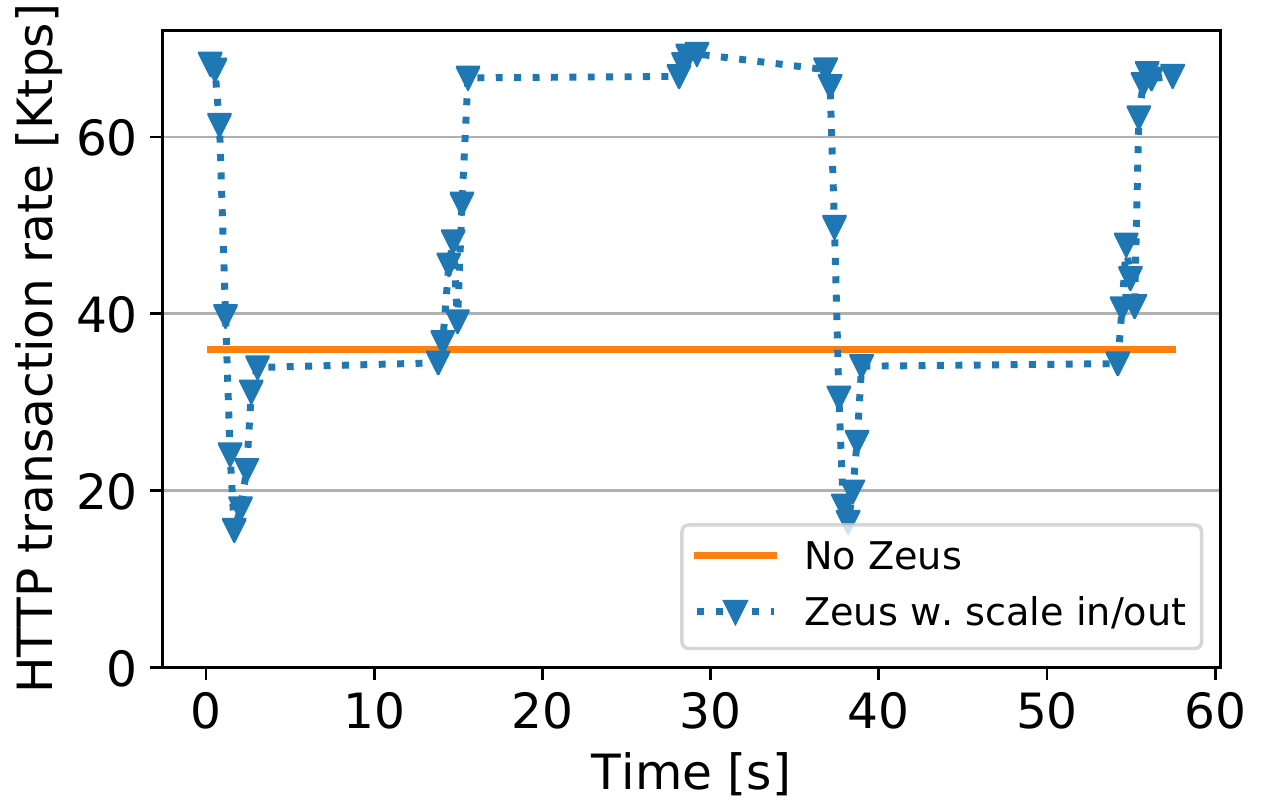}
    \vspace{-18pt}
    \caption{Nginx performance in a scale-in / scale-out scenario.}
    \label{fig:nginx_time}
\end{tabularx}

\vspace{-20pt}
\end{figure*}

\vspace{-5pt}
\subsection{Voter}

Voter is a benchmark that represents a real-time phone voting system~\cite{oltp2013bench}. 
Using three nodes, we simulate 20 contestants in a popularity show with 1M 
unique voters,
each identified by their phone number.
Each voter can vote for one contestant during one phone call and there is a limit how many times each voter may vote per unit of time. 
Therefore, each phone voting operation updates two objects: the total votes of a contestant and the voting history of the voter.

In this benchmark, we evaluate the ability of Zeus to move popular objects around, as discussed in Section~\ref{sec:background}. 
In the first experiment, we evaluate the performance of the ownership transfer protocol in isolation. 
We have 1M voters that generate 4M transactions per second (in comparison, \snewtext{E-store}~\cite{taft2014store} evaluates up to 200Ktps). 
At time 2s, we move all voter objects from node 1 to node 2, and at time 7s, we move them again to node 3. 
The results are shown in Figure~\ref{fig:voter_mov_1}.
We see that the full move takes 4s, implying that a single worker thread (out of ten) can move 25k objects per second. 

In the second experiment, we evaluate the performance of ownership transfers concurrently with transaction processing. 
We have 1 very popular contestant that has 100k voters voting for her, generating 700Ktps. 
All other voters vote for other contestants and generate about 5.3Mtps in aggregate. 
In this experiment, a single application and worker thread process the popular voter. 
As in the previous experiment, at times 2s, 6s and 10s, we start moving the object corresponding to the popular contestant to another node.
The results are shown in Figure~\ref{fig:voter_mov_2}.
We see that the single worker thread still performs 25k ownership requests per second (moving 100k objects in 4s) while at the same time the rest of the system completes 5.3Mtps.
This shows that the performance of ownership is not impacted by concurrent transactions. 

Figure~\ref{fig:cdf} shows the latency distribution of ownership transfer. 
This metric is important since an application thread is stalled during an ownership transfer, which allows easy porting of applications. 
We see that the mean latency and the $99.9^{\mathrm{th}}$ percentile are close during the first voter experiment; 17 and 36 $\mu$s, respectively. Under high load and while moving hot objects (during the second experiment) the mean latency is slightly higher at 29 $\mu$s, and the $99.9^{\mathrm{th}}$ percentile is 83 $\mu$s. This makes Zeus 3 times faster than Rocksteady\footnote{\newtext{Evaluated in similar setup with \CAP{DPDK} networking over 40Gb CX-3 \CAP{NIC}s.}}~\cite{Rocksteady} in the $99.9^{\mathrm{th}}$ percentile despite moving hot objects under load.

\subsection{Legacy applications}
One of the advantages of Zeus is that it is easy to port existing applications on it. 
Different applications assume different multi-threading or multi-process models, with different role for each thread (process). 
They also often take dependencies on various external libraries and \CAP{OS} calls. 
FaRM, FaSST and DrTM have to wait on each remote access.
To mitigate this latency, they
assume transaction multiplexing
% use 
via custom user-mode threading (e.g., co-routines or Boost user-threads in FaSST); however, this makes it difficult to integrate with many legacy applications.

As explained in Section~\ref{sec:design}, Zeus takes a different approach. 
Since most transactions are pipelined and do not block the application thread, there is no need to re-architect a legacy application. 
Zeus only blocks the application during the ownership requests, which are infrequent. 

In order to verify the claim about portability, we port and evaluate three existing applications on top of Zeus: the control plane of a cellular packet gateway, the \CAP{SCTP} transport protocol and an Nginx web server.

\vspace{2pt} \noindent {\bf Cellular packet gateway. }
Cellular packet gateway is a virtual network function in a cellular network that forwards all packets from mobile users. It has a control and data plane. 
The control plane performs service request and release operations, as described in the handover benchmark (but not the handovers themselves). 
Each of these operations is one transaction. 
We use the OpenEPCv8~\cite{OpenEPC} 4G implementation of the cellular core control plane. 
We remove the legacy datastore and instrument every access to use Zeus. 
We use a custom load generator to create test workloads with service and release requests. 
We test the gateway without any datastore (all data in local memory and no replication), using an off-the-shelf Redis datastore without replication, and Zeus. 

The results are shown in Figure~\ref{fig:spgw_thr}. 
Requests to Redis are remote and, due to the OpenEPC design, the application thread blocks on every request. 
This is why Redis performance is lower than 10Ktps even without replication, and illustrates the challenges due to blocking when porting existing applications. 
Zeus with a single active node (and 1 passive replica) is as fast as the gateway with local accesses and no replication. 
This is because the bottleneck is in parsing and processing the signalling messages, not in the datastore access. 
When we use both nodes as active (being each other replica), the throughput is 60\% higher. 
We are not able to scale beyond three nodes due to limitations of our signal generator, which cannot saturate more than two Zeus nodes.

\vspace{2pt} \noindent {\bf SCTP transport protocol. }
\CAP{SCTP} is commonly used in the cellular control plane to offer a degree of fault tolerance on network issues. 
For fault tolerance, \CAP{SCTP} natively supports multi-homing and is able to switch from one access network to another in case of a network failure, without dropping a connection. 
However, current \CAP{SCTP} implementations cannot survive a node failure as the connection state is not replicated.
If an \CAP{SCTP} connection fails, all active users drop calls.
Moreover, it is not easy to virtualize \CAP{SCTP} state as the protocol is originally implemented as a part of a Unix kernel. 

To demonstrate Zeus efficiency and the ability to support legacy applications, we port an implementation of \CAP{SCTP} protocol~\cite{usrsctp:2015} to Zeus and replicate all changes to the connection states.
We implement each packet transmission, reception and a timer event as a single transaction. 
Thus, any node failure will be perceived by the peers as a network loss, and dealt with by the protocol. 
\CAP{SCTP} uses standard \CAP{BSD} macros for basic data structures (\eg lists, hash tables) that are compatible with Zeus memory interfaces (described in Section~\ref{sec:system}).  
We are able to keep the original \CAP{SCTP} design (timer, \CAP{RX} and \CAP{TX} threads) as we do not have to deal with thread blocking.

We use a standard iperf3 client to generate a single \CAP{SCTP} flow to a Zeus server running \CAP{SCTP}. 
All state is replicated on another Zeus server. 
Figure~\ref{fig:sctp_thr} shows the throughput of the single flow for different packet sizes. 
For large packet sizes, Zeus is 40\% slower than vanilla \CAP{SCTP} with no modifications. 
This is because \CAP{SCTP} has a complex state that is modified for every packet and 6.8 KB of data has to be replicated 
(note that we have not spent any time optimizing state access and providing read-only accesses). 
The difference is higher for smaller packets because of the replication overhead. 
However, we argue that this is fine for the {\em control plane}, where the reliability is more important than speed.
We also note that Zeus pipelined transactions are important for the \CAP{SCTP} case with a few flows because many consecutive transactions access the same object and do not have to wait for the reliable commit of the previous transaction (\S~\ref{sec:tx-pipelining}).

\vskip 3pt \noindent {\bf Nginx web server. }
Finally, we evaluate the session persistence routing mode~\cite{nginx_session_persisten} of an Nginx web server on top of Zeus. 
In this mode, Nginx runs as an application-layer load balancer. 
It looks up a specific cookie in an \CAP{HTTP} request and chooses an end destination based on its value.
Session persistence is not available in the open source version of Nginx so we implement our own variant using the Zeus datastore. 
If the cookie is found in the replicated datastore, we route the request to the destination stored in the entry. 
If not, we randomly select one of the two \CAP{HTTP} back-end servers and store it to the datastore (replicated over two nodes).

A client creates a number of requests for a single small \CAP{HTTP} page.
Initially, all packets requests are processed by the same Nginx server node using a single core. 
We then emulate a scale-out and a scale-in by adding and removing another server node, and spreading the load across all available nodes.
The number of forwarded \CAP{HTTP} requests processed by Nginx is shown in Figure~\ref{fig:nginx_time}.
We see that the Nginx performance with Zeus is the same as without Zeus, showing that the bottleneck is in the application and not in the datastore. 
We also see that it seamlessly scales in and out as the number of servers change. 
Again, this illustrates an ease of portability of an existing legacy application to Zeus. 

\vspace{-3pt}
\section{Related work}
\vspace{-2pt}
\label{sec:related}

Recent works on in-memory distributed transactions present distributed commit protocols that leverage modern hardware to achieve good performance with strong consistency, but do not fully exploit locality~\cite{dragojevic2014farm,dragojevic2015no,kalia2016fasst,chen2016fast,wei2018deconstructing,sosp2015ramcloud}. Some systems expose object locality which allows programmers to implement locality-aware optimisations~\cite{dragojevic2014farm, sosp2007sinfonia}, 
% but object relocation is costly and is not transparent, as in Zeus.
but, unlike Zeus, object relocation is costly and burdens the programmer.

\newtext{
There are also works that mitigate the cost of distributed transactions but impose other constrains.}
For example, some mandate determinism~\cite{Lu:2020, Ren:2019, Le:2019, thomson2012calvin}, and are limited to non-interactive transactions that require the read/write sets of all transactions to be known prior to execution~\cite{ren2014evaluation}. 
\newtext{Others adopt epoch-based designs to amortize the cost of commit across several transactions~\cite{lu:2021, lu2018star, Crooks:18}.}
Contrary to those, Zeus enhances programmability and supports fully-general transactions \newtext{that need not wait the end of epochs to commit}.

% \beginbsec{Partitioning \& WAN}
Object partitioning has been used to improve performance of distributed transactions. Typically, objects are partitioned and migrated periodically to improve locality~\cite{curino2010schism, Psaroudakis:16, abebe2020dynamast, serafini2016clay, taft2014store, elmore2015squall, lee13asymmetric, Rocksteady}. 
In geo-distributed systems, object migration can significantly reduce \CAP{WAN} traffic~\cite{charapko2018adapting}. 
Facebook's Akkio~\cite{annamalai2018sharding} splits data in $\mu$-shards 
\newtext{which migrates across datacenters to leverage locality in workloads.
Similarly, \CAP{SLOG}~\cite{Ren:2019} deploys a periodic remastering scheme over a deterministic database to reduce across-datacenter round-trips, but mandates coordination within a datacenter.
Other works also exploit locality to reduce across-datacenter round-trips~\cite{taft2020cockroachdb, fan2019ocean, zhang2018building}.}
In contrast, Zeus infers locality and moves the object eagerly on the first access, \newtext{supports non-deterministic transactions,} and reduces coordination within the datacenter.

Zeus protocols bear similarity to 
cache coherence in multi-processor systems. Cache coherence protocols move the cache lines to the requesting node on access. Cache coherence protocols have been used to implement hardware transactions~\cite{htm}.
Zeus builds on ideas in Hermes~\cite{katsarakis2020hermes}, which adapted concepts from cache coherence and applied them to enforce strong consistency for replicated in-memory datastores. Hermes allows for local reads and fast reliable updates to individual objects from all replicas; however, it does not support multi-object reliable transactions or ownerships. 

% \beginbsec{Distributed shared memory}
Distributed shared memory 
(\CAP{DSM}) provides an abstraction of single shared memory space built on top of a collection of machines (e.g. ~\cite{munin,cashmere,treadmarks}). Similarly to Zeus, many \CAP{DSM}s use cache coherence protocols, moving data to the accessing node, but, unlike Zeus, most focus on single-object consistency. \newtext{A few support transactions (e.g., ~\cite{yu2018sundial, Cai:2018}) but relax consistency and/or forfeit availability for performance}. 

% \beginbsec{Ownership}
\newtext{Several works on software transactions have used owner\-ship-related ideas albeit on a single-node context~\cite{Marathe:08, Dice:06, Harris:14}.}
\newtext{
L-Store~\cite{lin2016towards} optimizes for locality using ownerships in a distributed local area setting, but only supports durable transactions (i.e., without replicas and availability). 
In contrast, Zeus enables strictly-serializable transactions and fast ownerships over a replicated deployment that facilitates availability and local read-only transactions from any replica.}

\newtext{
Akin to Zeus' local commit, \CAP{PWV}~\cite{Faleiro:17} enables early write visibility, as soon as a transaction executes all statements that could cause it to abort. 
However, transactions in \CAP{PWV} forfeit strictness and need determinism. In contrast, Zeus transactions exploit locality and afford strict serializability.}

An area that has looked into datastores with dynamic sharding are virtualized network functions. 
Several have built custom datastores to exploit locality (e.g.,~\cite{Woo:18, Sherry:15}), but they do not deliver on other desired requirements -- speed, availability or consistency. 
Others have forgone locality benefits~\cite{Kablan:2017, Khalid:19} and \newtext{rely on external datastores (e.g.,~\cite{ousterhout2015ramcloud})}.
\vspace{-4pt}
\section{Conclusion}
\vspace{-2pt}
\label{sec:conclusion}

Many real-world applications exhibit high access locality. 
Zeus leverages this to depart from the conventional distributed transaction design. 
Instead of executing a transaction across nodes, Zeus brings all objects to the same node and executes the transaction locally.
It does so via two new reliable protocols: one for fast localized transactions with replication, and one for efficient object ownership.  
Another benefit of Zeus is the ease of porting existing applications on top of it, as localized transactions can pipeline replication without blocking the application. 
Zeus is up to 2$\times$ faster than state-of-art systems on \CAP{TATP} benchmark and up to 40\% on Smallbank while using lower-end networking. 
It can move up to 250k objects per second per server and process millions of transactions per second. 
Zeus can run many industry standard applications without any re-architecting. 
We believe that Zeus can accelerate the uptake of reliable in-memory databases for a wide range of applications in the near future. 
% \begin{acks}

\vspace{4pt}
\beginbsec{Acknowledgments}
We thank our shepherd, Liuba Shrira, as well as Vitor Enes, Vasilis Gavrielatos, Vijay Nagarajan and our reviewers for their constructive comments and feedback.
This work is supported by the \CAP{EPSRC} grant \CAP{EP/L01503X/1} and by Microsoft Research via its PhD Scholarship Program.
% , as well as . 
% \end{acks}

% \begin{acks}
% We thank our shepherd, Liuba Shrira, and our anonymous reviewers for their constructive comments and feedback. 
% This work is supported by Microsoft Research through its PhD Scholarship Programme, as well as \CAP{EPSRC} grant \CAP{EP/L01503X/1}. 
% \end{acks}

%%% ACM Bibliography Style
\bibliographystyle{ACM-Reference-Format}
\bibliography{paper}

\end{document}